        \documentclass[structabstract,preprint]{aa}
        \usepackage{color}
        \usepackage{graphicx}
        \usepackage{txfonts}
        \usepackage{rotating}
        \usepackage{lscape}
        \usepackage{longtable,lscape}
        \usepackage{natbib}
        \bibpunct{(}{)}{;}{a}{}{,} 
	\newcommand{\mybf}{}
	\newcommand{\mybfnew}{}
	\newcommand{\mybfagain}{}
	\newcommand{\mybffinal}{}

\begin{document}
           \title{Proper motions of molecular hydrogen outflows in the $\rho$ Ophiuchi molecular cloud
        \thanks{Based on observations made with ESO New Technology Telescope at La Silla under programme ID 079.C- 0717(B), and on data obtained from the ESO Science Archive Facility.}
        }
   
        \author{M. Zhang\inst{1,2,3} \and
                       W. Brandner\inst{3} \and
                       H. Wang\inst{1} \and
                       M. Gennaro\inst{3,4} \and
                       A. Bik\inst{3} \and
                       Th. Henning\inst{3} \and
                       R. Gredel\inst{3} \and
                       M. Smith\inst{5} \and
                       Th. Stanke\inst{6}
        }

           \institute{Purple Mountain Observatory, \& Key Laboratory for Radio Astronomy, Chinese Academy of Sciences, Nanjing 210008, PR China\\
                      {\email{miaomiao@pmo.ac.cn}}
        \and {Graduate School of the Chinese Academy of Sciences, Beijing 100080, China}
        \and Max-Planck-Institut f\"ur Astronomie, K\"onigstuhl 17, 69117 Heidelberg, Germany
	\and Space Telescope Science Institute, 3700 San Martin Drive, Baltimore, MD 21218, USA
        \and Centre for Astrophysics and Planetary Science, The University of Kent, Canterbury CT2 7NH
        \and European Southern Observatory, Garching, Germany
        }

           \date{Received ---; accepted ---}


           \abstract
           {Proper motion measurements provide unique and powerful means to identify the driving sources of mass outflows, of particular importance
in regions with complex star formation activity and deeply embedded protostars. They also provide the necessary kinematic information to study the dynamics of mass outflows, the interaction between outflows and the ambient medium, and the evolution of outflows with age of the driving sources.}
           {We aim to take a census of \mybf{molecular hydrogen emission line objects (MHOs)} in the $\rho$ Ophiuchi molecular cloud and to make the first systematic proper motion measurements of \mybf{these objects} in this region. The driving sources are identified based on the measured proper motions, and the outflow properties are characterized. The relationship between outflow properties and the evolutionary stages of the driving sources are also investigated.}
           {Deep H$_2$ near-infrared imaging is performed to search for \mybf{molecular hydrogen emission line objects}. Multi-epoch data are used to derive the proper motions of \mybf{the features of these objects}, and the lengths and opening angles of the \mybf{molecular hydrogen} outflows.}
           {Our imaging covers an area of $\sim$ 0.11 deg$^2$ toward the L1688 core in the $\rho$ Ophiuchi molecular cloud. In total, \mybf{six new MHOs} are discovered, \mybf{32} previously known \mybf{MHOs} are detected, and the proper motions for 86 \mybf{features of the MHOs} are measured. The proper motions lie in the range of \mybf{14 to 247\,mas/yr}, corresponding to transversal velocities of \mybf{8 to 140\,km/s} with a \mybf{median} velocity of about \mybf{35\,km/s}. Based on morphology and proper motion measurements, \mybf{27 MHOs} are ascribed to \mybf{21} driving sources. The \mybf{molecular hydrogen} outflows have a \mybf{median} length of \mybf{$\sim$0.04 pc} and random orientations. We find no obvious correlation between H$_{2}$ jet length, jet opening angle, and the evolutionary stage of the driving sources as defined by their \mybf{spectral} indices. We find that the fraction of protostars (23\%) that drive \mybf{molecular hydrogen outflows} is similar to for Class II sources \mybf{(15\%)}. For most \mybf{molecular hydrogen} outflows, no obvious velocity variation along the outflow has been found.}
           {In Ophiuchus the frequency of occurrence  of \mybf{molecular hydrogen} outflows has no strong dependency on the evolutionary stage of the driving source during the evolution from the protostellar stage to the classical T Tauri stage.
} 
 
            \titlerunning{Proper motions of molecular hydrogen outflows in the $\rho$\,Oph cloud}
   \maketitle
%

\section{Introduction}
\mybf{Mass outflow  plays an important role during various stages of  star formation \citep{shu87,arce07,bally07}.} It is believed  that mass outflow \mybf{is a vital channel for the transfer} of excess angular momentum from the molecular cores to the ambient interstellar medium \citep{shang07}. Outflows are also good tracers of the young stellar objects (YSOs), in particular those deeply embedded in the molecular cloud cores. They can be studied in various regions of the electromagnetic spectrum. In the optical,
Herbig-Haro (HH) objects are the  manifestation of shock-ionized mass outflows, and trace either material ejected from the protostars directly or the shocked interstellar medium \citep{rb01,wang04,hhperseus,wang05,wang06,wang09}. The typical velocity of HH objects is up to several hundred km s$^{-1}$. At millimeter wavelengths, more than 400 CO molecular outflows have been detected \citep{wu04}. These CO outflows probe the swept-up or entrained medium in the outskirts away from the central jets, and have typical velocities of a few to ten km s$^{-1}$ \citep{bach96}.

In the near-infrared, molecular hydrogen emission lines, in particular the v~=~1-0~S(1) transition at the wavelength of 2.12 $\mu$m, are powerful tracers of shock-excitation. \citet{davis10} strictly defined molecular hydrogen emission-line objects (MHOs) as the near-infrared manifestations of outflows, and presented a comprehensive catalog of around 1000 MHOs. \mybf{Deep, wide-field near-infrared imaging with the combination of narrow-band H$_2$ filter and the corresponding filter for continuum emission such as a Ks filter has become a very efficient method to detect and identify MHOs \citep{stanke02,kha04,davis08,uwish2}.} Morphology, proper motion, as well as association with an HH object or CO outflow can all \mybf{provide information concerning} the location of the driving source. 
Once the relationship between the MHOs and their driving sources has been established, we can study the distribution of YSOs, the interaction of YSOs with the ambient interstellar medium, the star formation efficiency, and the evolution of the entire star forming region by analyzing the statistical characteristics of the MHOs. MHOs in several nearby star forming regions have been investigated using deep, sub-arcsecond-resolution, near-infrared surveys \citep{stanke02,davis08,davis09}.

The $\rho$ Ophiuchi molecular cloud is the nearest star forming region at a distance of about 120 pc \citep{lombardi08,loinard08}. \citet{lynds62} identified several dark nebulae in Ophiuchus by studying red and blue prints of the National Geographic-Palomar Observatory Sky Atlas. The large-scale structure of the $\rho$ Ophiuchi molecular cloud has been revealed by extensive molecular $^{13}$CO line mapping \citep{loren89a,loren89b}. Surveys for young stars in Ophiuchus include \citet{cas95}, \citet{kam97}, \citet{gro00}, \citet{ima02}, \citet{ozawa05}, and \citet{droxo10} in X-rays; \citet{wilking87} and \citet{wilking05} in the visual; \citet{gre92}, \citet{allen02}, and \citet{alv08} in the near-infrared; \citet{young86}, \citet{wilking01}, \citet{pad08}, \citet{evans09}, and \citet{irs10} in the mid- and far-infrared; and \citet{man98}, \citet{johnstone2000}, \citet{stanke06}, \citet{andr07}, and \citet{andr10} at sub-millimeter and millimeter wavelengths. By combining the results of the previous studies, \citet{book08} compiled a list of 316 young stars in L1688, the densest core of Ophiuchus. Using observations by the Spitzer space telescope, \citet{evans09} derived the most complete and unbiased YSO sample for the overall $\rho$ Ophiuchi star forming region. \mybf{\citet{evans09} investigated the overall evolution and estimated for the $\rho$ Ophiuchi cloud a star formation rate (SFR) per unit area of about 2.3 M$_{\sun}$ Myr$^{-1}$ pc$^{-2}$. This is lower than the SFR in the Serpens cloud (3.2 M$_{\sun}$ Myr$^{-1}$ pc$^{-2}$) but higher than the SFR in the Perseus cloud (1.3 M$_{\sun}$ Myr$^{-1}$ pc$^{-2}$) \citep{evans09}.}

Outflows in Ophiuchus have been investigated in some studies at different wavelengths. Close to 50 Herbig-Haro (HH) objects \citep{wilking97,gome98,wu02,phe04} and more than 15 high-velocity CO molecular outflows \citep{andre93,dent95,bon96,wu04,bus07,naka11} have been identified. Using mid-infrared Spitzer/IRAC observations, \citet{zw09} detected 44 extended green object (EGO) outflow features. Based on near-infrared data, \citet{davis10} list $\sim$47 MHOs consisting of $\sim$119 H$_{2}$ 2.12 $\mu$m emission features in the $\rho$ Ophiuchi molecular cloud. While observations with small angular coverage detected some individual MHOs \citep{grosso01,yba06}, the majority of MHOs in Ophiuchus have been detected by two unbiased surveys \citep{gome03,kha04}. We observed a subset of these MHOs, aiming at the second epoch images to measure the proper motions (PMs) of \mybf{the MHO features}.

Compared with outflow morphology, motions of outflow features are more robust evidence for the identification of the driving sources of outflows. The directions of motion vectors indicate the location of the driving sources of the outflow features. L1688, the densest region in Ophiuchus, includes hundreds of young stars and several tens of \mybf{outflow features} within an area of less than one square degree. The relationship between the young stars and \mybf{the outflow features} is very complex in this region. Previous studies associated the \mybf{outflow features} with the young stars just based on the morphologies \citep{phe04,gome03,kha04}, which have large uncertainties. \mybf{The kinematic information} of the MHOs, including proper motions and radial velocities, will offer reliable evidence for the identification of the driving sources in this region.

This work is part of a project aimed at understanding the dynamical characteristics of outflows in Ophiuchus. The objectives of the project are to measure the radial velocities of jets with VLT/CRIRES, and to derive the proper motions of jets based on multi-epoch astrometry with NTT/SofI. This then in turn should enable us to derive the flow orientation, trace the driving sources of the jets, and based on the 3D velocity information and spectral line features to compare observed jet properties with models of shock physics. As a first step, we here present the proper motion measurements of MHO features in the $\rho$ Ophiuchi molecular cloud. The radial velocity measurements of MHO features with VLT/CRIRES in Ophiuchus will be published in another paper (Zhang et al. 2013, in preparation).

\section{Observations and data reduction}

\subsection{Near-IR imaging}\label{data}

The observations were conducted in service mode between July and September 2007 (079.C- 0717(B), PI: M.D. Smith) with the ESO NTT near-infrared spectrograph and imaging camera SofI \citep{sofi}, covering in total an area of $\sim$ 0.11 deg$^2$. Figure \ref{figcover} shows the coverage of the observations with the gray filled boxes in the left panel. SofI is equipped with a Hawaii HgCdTe 1024$\times$1024 array, yielding a field of view of $\sim$4\farcm9$\times$4\farcm9 with a plate scale of 0\farcs288 pixel$^{-1}$ in its large field (LF) imaging mode. In total, 16 fields were observed in K$_{s}$ and H$_{2}$ band. The broadband K$_{s}$ filter has a central wavelength of 2.162 $\mu$m and a bandwidth of 0.275 $\mu$m. The narrowband H$_{2}$ filter has a central wavelength of 2.124 $\mu$m and a bandwidth of 0.028 $\mu$m. Using dithered observations, each field was typically imaged seven times with the individual exposure time of 18s in K$_{s}$ band and of 180s or 240s in H$_{2}$ band. Table \ref{obslog} lists the center coordinates, the total integration time, and the average airmass for each field.

\begin{table}[htb]
\centering
\footnotesize
\caption[]{\centering Observing log}
         \label{obslog}
\begin{tabular}{cccccc}
\hline\hline
 \multicolumn{2}{c}{Center of field} & \multicolumn{2}{c}{Exposure time} &\multicolumn{2}{c}{Airmass}   \\ \hline
    RA& Dec.   &  H$_{2}$ & Ks& H$_{2}$ & Ks \\
(J2000) & (J2000) & (s) & (s) & & \\ \hline
  16 26 16.0  &-24 24 30 & 1680 & 126 & 1.02 & 1.01 \\
 16 26 21.0  &-24 27 55 & 1680 & 126 & 1.05 & 1.04 \\
 16 26 33.0  &-24 25 20 & 1680 & 126 & 1.10 & 1.09 \\
 16 26 53.9  &-24 37 33 & 1260 & 126 & 1.18 & 1.17 \\
 16 27 03.3  &-24 47 03 & 1260 & 126 & 1.27 & 1.26 \\
 16 27 08.8  &-24 33 43 & 1680 & 144 & 1.49 & 1.47 \\
 16 27 11.8  &-24 43 59 & 1260 & 126 & 1.75 & 1.72 \\
 16 27 22.4  &-24 49 13 & 1260 & 126 & 1.08 & 1.07 \\
 16 27 24.3  &-24 28 17 & 1680 & 198 & 1.14 & 1.12 \\
 16 27 34.1  &-24 39 18 & 1260 & 126 & 1.64 & 1.61 \\
 16 27 43.0  &-24 40 47 & 1260 & 126 & 1.24 & 1.23 \\
 16 27 45.5  &-24 45 42 & 1260 & 126 & 1.37 & 1.35 \\
 16 27 47.7  &-24 31 50 & 1260 & 126 & 1.89 & 1.85 \\
 16 28 01.6  &-24 31 02 & 1260 & 126 & 1.53 & 1.51 \\
 16 28 10.2  &-24 16 01 & 1260 & 126 & 1.92 & 1.88 \\
 16 28 20.3  &-24 37 07 & 1260 & 126 & 1.37 & 1.36 \\
\hline
\end{tabular}

\end{table}

We processed the data using the IRAF packages. After subtracting a median dark frame, each image was divided by a normalized flat-field for each filter. We then corrected the illumination effect and masked the bad pixels using the calibration data available at ESO's website\footnote{http://www.eso.org/sci/facilities/lasilla/instruments/sofi/tools\\
/reduction/index.html}, and then subtracted the sky frames, which had been generated from a median-filtered set of the flattened frames in Ks and H$_{2}$ bands. For the SofI imaging mode, a bright source on the array produces a ``ghost'' that affects all the lines where the source is and all the corresponding lines in the other half of the detector, a instrumental feature called ``Inter-quadrant Row Cross Talk" (details can be found in SofI User's Manual\footnote{http://www.eso.org/sci/facilities/lasilla/instruments/sofi/doc/manual\\
/sofiman\_2p10.pdf}).
We also corrected this instrumental effect with the IRAF script from SofI website\footnote{http://www.eso.org/lasilla/instruments/sofi/tools/reduction\\
/sofi\_scripts/}. In order to get the accurate astrometry, we also removed the geometrical distortion on the dithered frames using the geometric distortion solution.
Finally we shifted and combined the dithered frames in each filter for each field after excluding several images with bad quality.

\subsection{Proper motion measurements}\label{pm}
\begin{figure*}[htb]
\centering
\includegraphics[width=\textwidth]{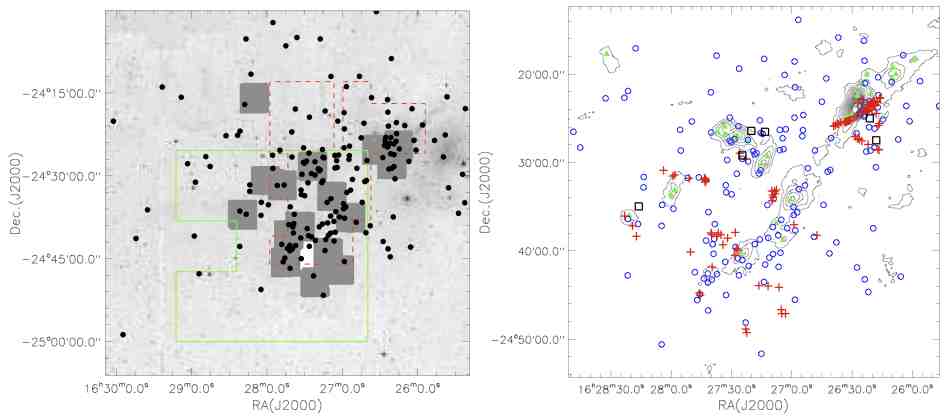}
\caption{Coverage of our data. In \textit{the left panel}, the underlying image is 2MASS image in Ks-band. The gray filled boxes show the coverage of \textit{SofI} (2007, PID: 079.C- 0717(B)) observations. The red dashed and green solid polygons outline the coverage of the \textit{SofI} 2000 (PID: 65.I- 0576(A)) and 2001 (PID: 67.C- 0284(A)) observations, respectively. The YSOs identified by Spitzer photometry \citep{evans09} are marked with filled circles. In \textit{the right panel}, the gray contours represent the continuum emission at 1120 $\mu$m from the \textit{COMPLETE} project \citep{com06} \ with the contour levels from 80 to 2000 mJy/beam and a step of 100 mJy/beam. The green filled triangles show the positions of the 1100 $\mu$m \ millimeter sources identified by \citet{young06}. The newly detected MHO features are marked with black open squares and the previously known MHO features are labeled with red pluses. The YSOs identified by Spitzer photometry are marked with blue circles.}
\label{figcover}
\end{figure*}

Archival SofI images ((PID: 65.I- 0576(A) and 67.C- 0284(A)) obtained in July 2000 and August 2001, and published by \citet{gome03,kha04} served as the first epoch data. The newly-acquired SofI images provided the second epoch to measure the proper motions of the H$_2$ emission-line features in MHOs. We reprocessed the data using an IDL pipeline which was improved by \citet{mario12} according to the process of data reduction described in Section. \ref{data}. Figure \ref{figcover} shows the field coverage. The gray filled boxes show the coverage of the \textit{SofI} (2007, PID: 079.C- 0717(B)) observations\mybf{, while} the \mybf{red dashed and green solid} polygons outline the coverage of the \textit{SofI} 2000 (PID: 65.I- 0576(A)) and 2001 (PID: 67.C- 0284(A)) observations, respectively.

In order to obtain accurate proper motion measurements, we \mybf{firstly corrected for seeing}. For each image in the image pair (first epoch image and corresponding second epoch image), the average full width at half maximum (FWHM) of all detected stars is used to represent the seeing condition. Then the image, which has the smallest average FWHM is convolved with a Gaussian core so that two images in each image pair have the same FWHM. After convolution, we align and resample the images in order to match the \mybf{image pairs} pixel to pixel. The astrometric calibration was performed using the IRAF packages. Using stars in common between the different data sets, we registered the first epoch SofI images to the 2MASS Point Source Catalog, and then registered each second epoch SofI image to the corresponding first epoch image. \mybf{The root-mean-square (rms) values for the registration residuals of all fields were below 0.1\arcsec~($\sim$1/3 pixel).}

\mybf{The proper motions of the MHO features are measured with a method similar to the one suggested by \citet{mcg07} and \citet{davis09}.} For each \mybf{MHO} feature, we defined manually a polygon aperture based on the surface brightness distribution to estimate the flux and morphology of the feature. Then the first epoch image was shifted relative to the second epoch image in steps of 0.25 pixels \mybf{about an 8 $\times$ 8 pixel grid}. At each position the flux of the features in the polygon are used to calculate the cross-correlation coefficient. This process yields a set of correlation coefficients as a function of shifts ($ccf = f(shiftx, shifty)$). Obviously, the shift corresponding to the maximum correlation coefficient can be obtained as the shift of each feature between two epochs. \mybf{Note that we determined the position of the maximum correlation coefficient by calculating an intensity weighted center using pixels in the correlation coefficient image around the peak value.} Finally, we subtracted the apparent proper motions of the adjacent background field stars from this shift in order to minimize any systematic effects such as the problems with alignment, resampling, or astrometry.
As for the method described in detail by \citet{davis09}, the errors of PMs of flow features are mainly from the random motions of the field stars. 

The discussion above assumes that the brightnesses and morphologies of the \mybf{MHO} features do not change between the first and second-epoch observations: for many features this is not the case. In fact, many factors such as signal-noise-ratio, shape, flux variability of the \mybf{MHO} features, and the gradient in the background emission also contribute to the errors of the PM measurements, but these factors are very hard to quantify. Therefore, we checked every feature and introduced a parameter called ``Status" to estimate the reliability of the PM calculations. A PM with status 0 is reliable, while a PM with status 1 or 2 is uncertain. 

\mybf{The proper motions of MHO features obtained through the above process are relative to the background field stars. \citet{makarov07} investigated 58 probable members of the association of pre-main-sequence stars around the filamentary $\rho$ Ophiuchi cloud using astrometric proper motions from the UCAC-2 catalog and the convergent point method. They found that the selected probable members have similar proper motions with an average of roughly $(\mu_{\alpha} cos\delta, \mu_{\delta}) = (-11.4, -24.5)$ mas yr$^{-1}$. The bulk of the standard proper motion errors are in the range 1-12 mas yr$^{-1}$ per coordinate direction. \citet{mamajek08} also investigated 53 candidate members in L1688 based on the Tycho-2, UCAC2 and SPM2.0 catalogs. They calculated the mean proper motion \mybfnew{for members in the group} with several estimators which are fairly insensitive to outlying points and finally obtained the mean proper motion of $(\mu_{\alpha} cos\delta, \mu_{\delta}) = (-10, -27)$ mas yr$^{-1}$ with the total uncertainty of (2, 2) mas yr$^{-1}$. We adopted the mean proper motion from \citet{mamajek08} to represent the systematic proper motion of the members in L1688. We subtracted this systematic proper motion from all the PM measurements of the MHO features with the aim of getting the proper motions of the MHO features with respect to the $\rho$ Ophiuchi molecular cloud.} A complete list of proper motions of MHO features in $\rho$ Ophiuchi is provided in the Appendix.

\subsection{Flux estimation of MHO features}\label{flux}

\mybf{Areal photometry was performed to estimate the flux of MHO features. We firstly defined a polygon aperture based on the morphology and surface brightness distribution of each MHO feature. A 10 pixel wide region of pixels surrounding the polygon aperture is used to estimate the sky background. The polygon apertures and the surrounding annulus are applied to the MHO features on the continuum-subtracted H$_2$ images to estimate the fluxes of H$_2$ 1-0 S(1) emission line.}

\mybf{The flux calibration was obtained via the 2MASS catalog. We used Sextractor \citep{sex} to do the photometry for the point sources on the H$_2$ images. The zero-point flux was derived through the comparison of the integrated counts with the cataloged flux of the point sources. We applied the zero-point flux to the integrated counts of the MHO features with the aim to convert the counts of MHO features to the flux unit.}

\begin{figure}[htb]
\centering
\includegraphics[scale=0.5]{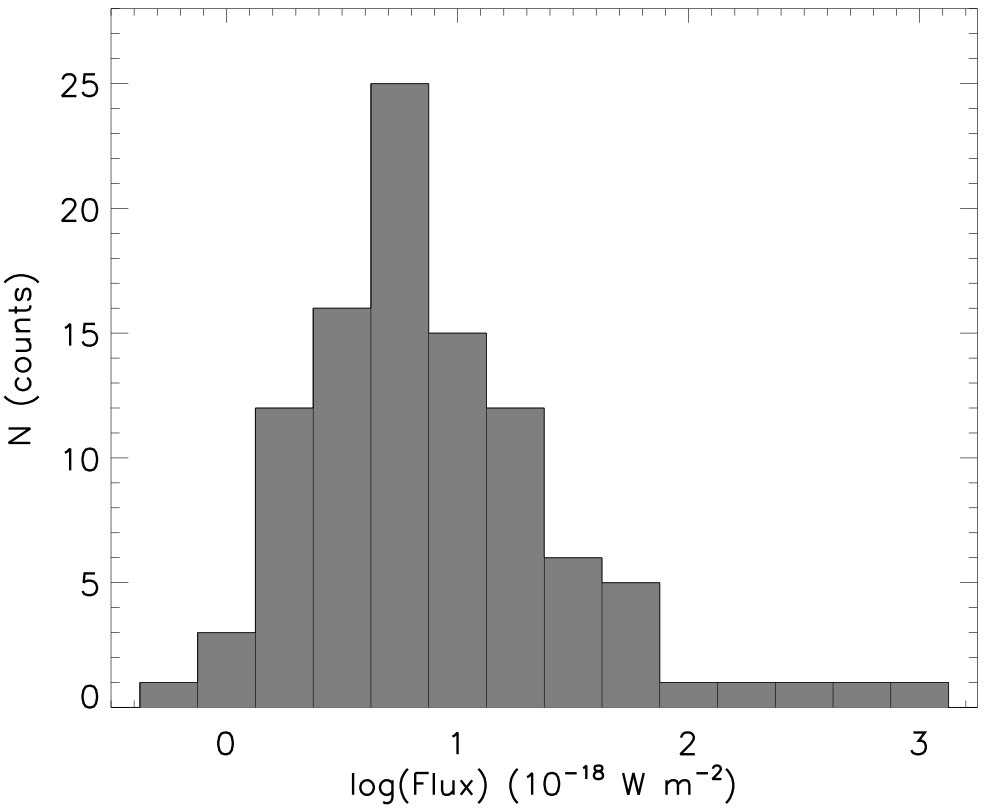}
\caption{Histogram of integrated 1-0 S(1) fluxes for the detected MHO features.}
\label{figflux}
\end{figure}

\mybf{Figure.~\ref{figflux} shows the distribution of the fluxes of the MHO features. The median of the fluxes of MHO features is about 6$\times$10$^{-18}$ W m$^{-2}$. The minimum of the fluxes we detected is about 6$\times$10$^{-19}$ W m$^{-2}$.}

\section{Results}
\subsection{The detected MHOs in Ophiuchus}
We performed our survey towards the L1688 core, which is the densest region in the $\rho$ Ophiuchi molecular cloud, covering an area of about 0.1 deg$^2$. We detected \mybf{six} new MHOs and \mybf{32} known MHOs, with in total \mybf{107} MHO features. Table \ref{newmho} lists the position and morphology of \mybf{six} newly detected MHOs and figure \ref{figcover} shows the distribution of all 107 MHO features.

\mybf{Note the difference between ``MHO" and ``MHO feature". An MHO feature is an H$_2$ emission feature which can not be resolved to more sub-structures in H$_2$ images. An MHO consists of one or more MHO features. Therefore, we detected 38 MHOs which consist of 107 MHO features and the PM measurements and photometry mentioned above are calculated for the MHO features.}

MHO 2149, MHO 2154, \mybf{and MHO 2155} are faint features which are only detected in our deep \textit{SofI} (2007) images. MHO 2151-2153 are visible in two epoch images. \mybf{We reprocessed the \textit{SofI} (2001) data and detected MHO2151-2153, which were not identified by \citet{kha04} in the first epoch images.} Therefore, we also obtained the proper motions of MHO 2151-2153. The details of all detected MHOs in Ophiuchus (including names, positions, proper motions and the notes on individual objects) can be found in the Appendix.

\begin{table*}[htb]
\centering
\footnotesize
\caption[]{\centering Newly detected \mybf{MHOs} in Ophiuchus}
         \label{newmho}
\begin{tabular}{lccl}
\hline\hline
MHO names\tablefootmark{a}&RA & Dec. & Comments   \\
    &      (J2000) & (J2000)&  \\ \hline
MHO 2149       &+16 26 17.9         &-24 27 29.4         &faint knot                                        \\
MHO 2151       &+16 26 21.2         &-24 25 00.3         &three faint knots in a diffuse nebula            \\
MHO 2152       &+16 27 13.3         &-24 26 31.9         &knot with a faint diffuse nebula \\
MHO 2153       &+16 27 20.1         &-24 26 25.5         &knot                                              \\
MHO 2154       &+16 28 16.1         &-24 34 58.2         &faint diffuse  nebula\\
MHO 2155       &+16 27 24.6         &-24 29 12.0         &faint elongated knot\\ 
\hline
\end{tabular}
\tablefoot{
\tablefoottext{a}{These names are provided by C.J.\ Davis and will be included in the online MHO catalog (http://www.astro.ljmu.ac.uk/MHCat/). \mybf{Please note that there is no MHO 2150 in this table. Actually, MHO 2150 has been detected by \citet[][see Fig. 12 in their paper]{cog06}, but somehow not been included in the MHO catalog before. We thus associated this object with MHO 2150 and included it in Table. \ref{mhopm} as a previously known MHO.}}
}
\end{table*}

\subsection{The proper motions of MHO features}
We obtained the proper motion measurements for 86 MHO features in Ophiuchus. Table~\ref{mhopm} in the Appendix lists the proper motion of each MHO feature. Here we  present some statistics of these PM measurements.

\begin{figure}[htb]
\centering
\includegraphics[scale=0.5]{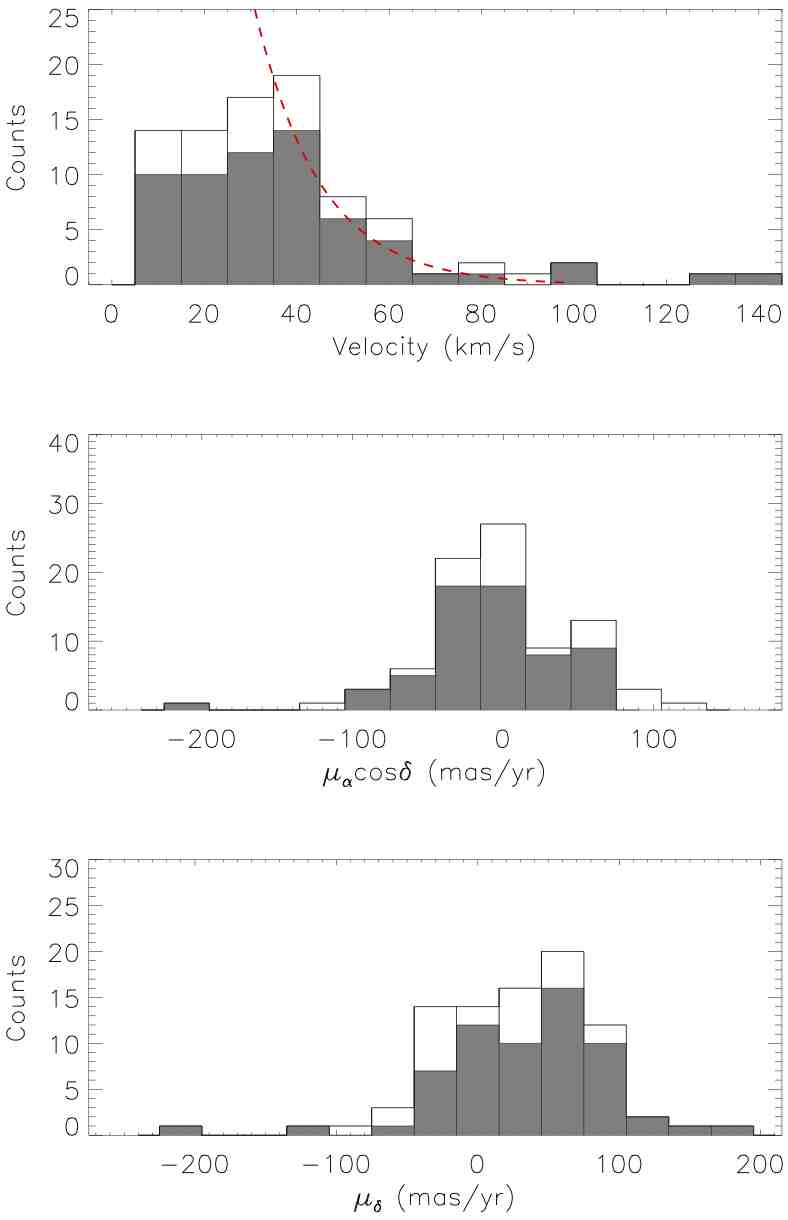}
\caption{\textit{Top}: distribution of tangential velocities of MHO features (in 10 km s$^{-1}$ \ bins); \textit{middle}: the distribution of $\mu_{\alpha}cos\delta$ (in 30 mas yr$^{-1}$ \ bins); \textit{bottom}: the distribution of $\mu_{\delta}$ (in 30 mas yr$^{-1}$ \ bins). The open columns are for all PM measurements of \mybf{MHO} features in Ophiuchus and the gray filled columns are for the PM measurements with status of `0' (reliable measurements) of \mybf{MHO} features. The red dashed line in top panel shows the fitting with the exponential function for the MHO features with the reliable tangential velocities greater than 40 km s$^{-1}$.}
\label{figpmsta}
\end{figure}

We found that the proper motions of \mybf{MHO} features are in the range of \mybf{14 to 247} mas yr$^{-1}$, corresponding to transversal velocities of \mybf{8 to 140} km s$^{-1}$ if assuming that all of \mybf{MHO} features are located at a distance of 119 pc \citep{lombardi08}. Figure. \ref{figpmsta} shows the distribution of tangential velocities (top panel), $\mu_{\alpha}cos\delta$ (middle panel) and $\mu_{\delta}$ (bottom panel) for all PM measurements of \mybf{MHO} features with the open columns and for reliable PM measurements (Status ``0") with the gray filled columns. \mybf{The median of the distribution of tangential velocities is 34.5 km s$^{-1}$ for all PM measurements of MHO features and 34.7 km s$^{-1}$ for the reliable PM measurements. The peak of the distribution of tangential velocities is in the range of 30-40 km s$^{-1}$ with the variation in the bin size. Moreover, the MHO features with tangential velocities $>$ 40 km s$^{-1}$ show an exponentially decaying frequency. Actually, we fitted the tangential velocity distribution of reliable PM measurements (marked with status 0) with the value of $>$ 40 km s$^{-1}$ using the power-law function and the exponential function, individually. The fitting with the exponential function returned the smaller chi-square value. Therefore, we think that the distribution of tangential velocities with the value $>$ 40 km s$^{-1}$ is closer to the exponential distribution. \mybffinal{The top panel of Figure. \ref{figpmsta} shows the fitting result with the red dashed curve. The counts of MHO features is related to the tangential velocities of MHO features in the way of Counts $\propto$ e$^{(-0.071\pm 0.011)\times velocity(km/s)}$.}}

\citet{davis09} measured proper motions for 147 H$_2$ features in Orion A. \mybf{The median of the distribution of tangential velocities is 70.8 km s$^{-1}$ for all PM measurements of MHO features and 75.3 km s$^{-1}$ for the reliable PM measurements of MHO features in Orion. The peak of the distribution of reliable PM measurements of these MHO features in Orion is in the range of 60-80 km s$^{-1}$. \mybffinal{The distribution of tangential velocities with the value $>$ 70 km s$^{-1}$ also shows an exponential distribution in the way of Counts $\propto$ e$^{(-0.028\pm 0.004)\times velocity(km/s)}$.}} 

\mybf{We also note that there are drops in number at low velocity in the distributions of tangential velocities of MHO features in Ophiuchus and in Orion. These drops should result from the incompleteness of the PM measurements. For the low surface brightness MHO features and the slow moving MHO features, we cannot obtain reliable proper motion measurements. Therefore, these drops are due to a bias. More sensitive observations with higher angular resolution are required to construct the correct distribution in the low velocity range.}

\subsection{The driving sources of MHOs in Ophiuchus}
The most probable driving sources of MHOs are identified based on the outflow morphology and proper motion analysis. In total, we associated \mybf{27} MHOs with \mybf{21} YSOs identified by \citet{evans09}. Table \ref{promho} lists these \mybf{27} MHOs and their likely driving sources. The driving sources of the remaining \mybf{eleven} MHOs can not be identified because of the uncertainties in their proper motions. Figure \ref{figstat} shows the distribution of the \mybf{21} driving sources of MHOs in Ophiuchus.

\mybfnew{Note the difference between the number of MHOs (27) and the number of molecular hydrogen outflows (21). A molecular hydrogen outflow may consist of two or more MHOs that are associated with the same driving source.}

Previous studies \citep{gome03,kha04} also associated \mybf{MHOs} in Ophiuchus with some young stars, but their identifications are only based on MHO morphology and there are large uncertainties. Our proper motion measurements supply more robust evidence for the driving source identifications of MHOs in Ophiuchus, especially for the complicated regions that have a large number of young stars around the MHOs. For example, in the region with VLA1623 (see Figure. \ref{figA1003main}), there are several tens of \mybf{MHO} features and tens of young stars. Previous studies \citep{dent95,gome03} suggested that most of the MHO features in this region should belong to the outflow which is driven by VLA 1623 based on the association between the \mybf{MHO} features and the CO outflow driven by VLA 1623. But they cannot point out which features are exactly excited by VLA 1623 and which are not, especially for the features near the young star GSS 30. We derived reliable PM measurements for most of \mybf{MHO} features in the region of VLA 1623 (also see Figure. \ref{figA1003main}). We found that most of the PM vectors trace back to VLA 1623, which confirms that these \mybf{MHO} features are driven by VLA 1623. We also found that there is no obvious evidence for the assumption that some \mybf{MHO} features in this region are excited by GSS 30.

\mybf{Among 24 previously known MHOs which are associated with the driving sources in our paper based on the morphologies and the PM measurements, there are 13 MHOs ($\sim$54\%) whose driving sources are in agreement with the previous identifications while only 5 MHOs ($\sim$21\%) are associated with the different driving sources from the previous identifications. Note that there are 4 MHOs ($\sim$17\%) which were associated with several possible driving source candidates in previous studies. Now we can provide a robust identification for their driving sources based on our PM measurements. Note that there are two previously known MHOs (MHO 2103 and MHO 2104) for which we did not obtain the PM measurements because of the lack of the first epoch \textit{SofI} data.}

\citet{naka11} have published the results of CO mapping observations toward the L1688 core in Ophiuchus. They identified five CO molecular outflows based on the CO (J = 3-2) and CO (J = 1-0) emission maps. We also checked for the associations between MHOs and CO outflows. Table \ref{promho} lists the results. We found that \mybf{two} of the driving sources of MHOs also drive the CO outflows. \mybf{The MHOs driven by VLA 1623-243 are well associated with the CO outflows detected in the CO (J = 3-2) emission map. Many MHO features that are driven by VLA 1623-243 correspond to the red-shifted lobe of the CO outflow.} \citet{naka11} also detected a weak CO outflow on their CO (J = 1-0) map and they associated this CO outflow with the YSO IRS 44. This CO outflow was firstly identified by \citet{tere89} by means of CO (J = 1-0) interferometric maps using the Owens Valley Millimeter Interferometer. We have compared the position and the morphology of their CO outflow \citep[see][Fig. 2b]{tere89} with our MHOs and we suggest to associate this CO outflow with YLW 16 -- which also drives two MHOs -- rather than IRS 44. We also detected \mybf{an MHO, MHO 2153}, corresponding to the blue-shifted lobe of the CO outflow that is driven by Elias 2-32. \mybf{However, according to the proper motions we obtained for MHO 2153, we associate MHO 2153 with [GY92] 239 rather than with Elias 2-32.} For the CO outflow driven by Elias 2-29, we can not detect any MHO counterparts as our H$_{2}$ SofI (2007) map does not cover the region around Elias 2-29.


\citet{johnstone2000} have published the results from a survey of the central 700 arcmin$^2$ region of the $\rho$ Ophiuchi molecular cloud at 850 $\mu$m using the submillimeter common-user bolometer array (SCUBA) on the James Clerk Maxwell Telescope and they identified 55 cores on their 850 $\mu$m map. Moreover, \citet{young06} also published the results from large-scale millimeter continuum map at 1.1 mm of the $\rho$ Ophiuchi molecular cloud with Bolocam on the Caltech Submillimeter Observatory (CSO) and they detected 44 definite sources. \citet{evans09} correlated these dust cores with their YSO sample in order to get the complete spectral energy distributions (SEDs). Among our \mybf{21} flow sources, \mybf{four sources (19\%)} have 850 $\mu$m core counterparts and four sources \mybf{(19\%)} have 1.1 mm core counterparts. In fact, there are about 184 YSOs in the L1688 core region, of which 23 YSO ($\sim$12.5\%) are associated with 850 $\mu$m dust cores, and 22 YSOs ($\sim$ 12\%) have 1.1 mm core counterparts. \mybf{YSOs driving molecular hydrogen outflows tend to have a higher degree of association with millimeter/sub-millimeter cores than YSOs not driving molecular hydrogen outflows.}

\subsection{Outflow statistics}
Based on the positions and proper motions of the MHO features, we calculated the jet length (\textit{L}), jet opening angle (\textit{$\theta$}), jet position angle (PA), variation in the velocity along the outflow (VV), and dynamic ages (DA) for the \mybf{molecular hydrogen outflows}. \mybf{We defined some parameters (\textit{L}, \textit{$\theta$}, PA) according to the suggestions from \citet{davis09}.}

Jet opening angle (\textit{$\theta$}) is measured from a cone with the smallest vertex angle centered on the driving source that includes all MHO features in the outflow. The opening angle can not be calculated if there is only one feature in the outflow. Jet position angle (PA) is measured east to north as computed from the bisector of the opening angle. We selected the angle smaller than 180 degrees as the position angle for the bipolar outflows. Jet length (\textit{L}) is defined as the distance from the MHO feature to the source projected to the bisector of the opening angle in cases where more than one feature is identified in the outflow. In case where there is only one feature identified in the outflow, this length is just the distance from the source to the feature. Variation in velocity (VV) along the outflow is calculated through fitting the tangential velocity of MHO features vs. the distance from MHO features to the driving source using a linear least square method. \mybf{Note that we calculate the VV of each lobe for the bipolar outflows.} VV equals 1 means that the MHO features will speed up by 1 km s$^{-1}$ after they propagate 100\,AU along the outflow \mybf{if assuming that the outflows are located at the distance of 119 pc}. The dynamic age (DA) of the outflow is the maximum dynamic age of the features with proper motions. Table \ref{promho} lists these parameters for each outflow. Note that these parameters are only calculated with MHOs (not considering the information provided by HH objects and CO outflows) and uncorrected for inclination to the line of sight. 

\begin{figure}[htb]
\centering
\includegraphics[scale=0.5]{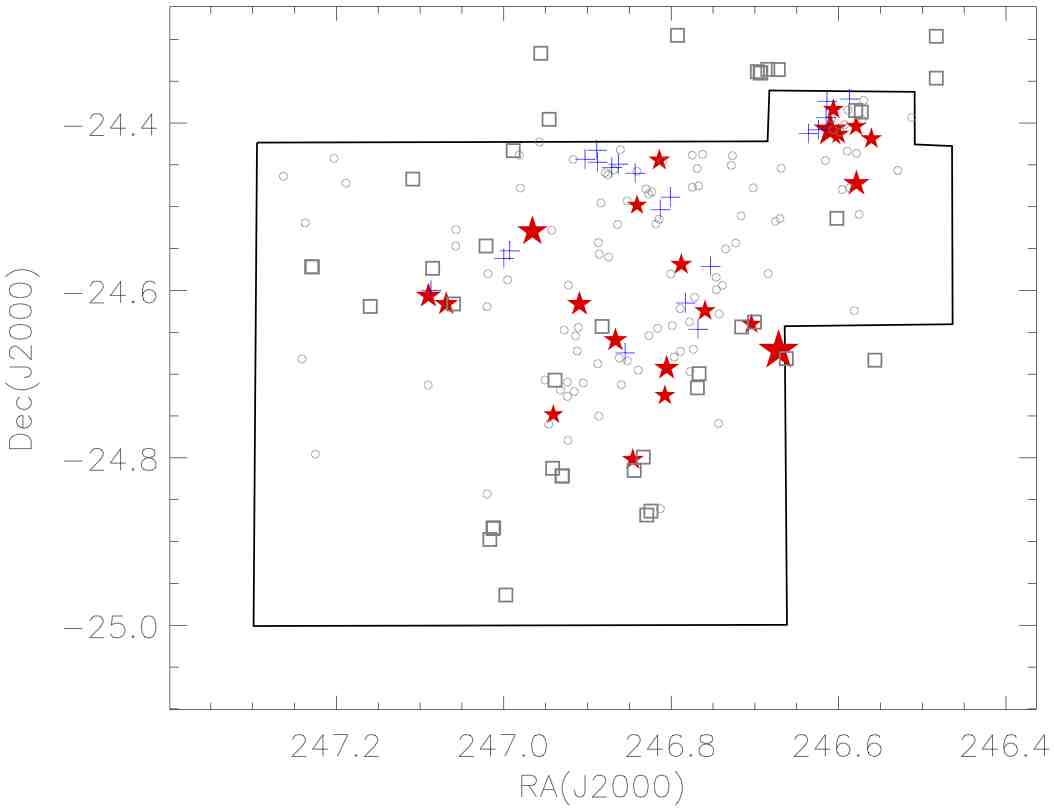}
\caption{Distribution of outflows in Ophiuchus. The driving sources of outflows identified in this paper are marked with red filled pentagrams whose sizes are proportional to the length of the outflow. The YSOs identified by \citet{evans09} are labeled with open gray circles and the millimeter sources identified by \citet{young06} are marked with blue pluses. The open gray squares show the distribution of known Herbig-Haro objects (from the \textit{SIMBAD} database). The region that we have selected for outflow statistical study is outlined with a black solid polygon. Note that for the YSOs and millimeter sources only those that are located inside this polygon are plotted.}
\label{figstat}
\end{figure}

As our data cover only part of Ophiuchus, we selected the region outlined in Figure \ref{figstat} with a black solid polygon according to the coverage of unbiased surveys conducted by \citet{kha04}, \citet{gome03} and the coverage of our \textit{SofI} (2007) data. In this region, our MHO sample is complete. This area covers the majority of the L1688 core, and our follow-up statistics are all restricted to this area.

\mybf{In this area, there are 123 YSOs \citep{evans09}, including 52 protostars (-0.5$<$$\alpha$$<$3.0) and 89 disk-excess sources (-2.5$<$$\alpha$$<$0.0). Of these 52 protostars, 10 (19\%) sources drive outflows. This fraction is lower than the fraction ($\sim$30\%) of protostars with outflows in Orion A \citep{davis09}. \citet{davis09} identified 116 jets in Orion A and most of their driving sources are protostars. However, of our 21 sources with outflows, 10 are protostars and 12 are disc-excess sources. The criterion of the classification for protostars and disk-excess sources is from \citet{davis09} and note that there is an overlap between the range of $\alpha$ for protostars and for disc-excess sources.}

\begin{figure}[htb]
\centering
\includegraphics[scale=0.5]{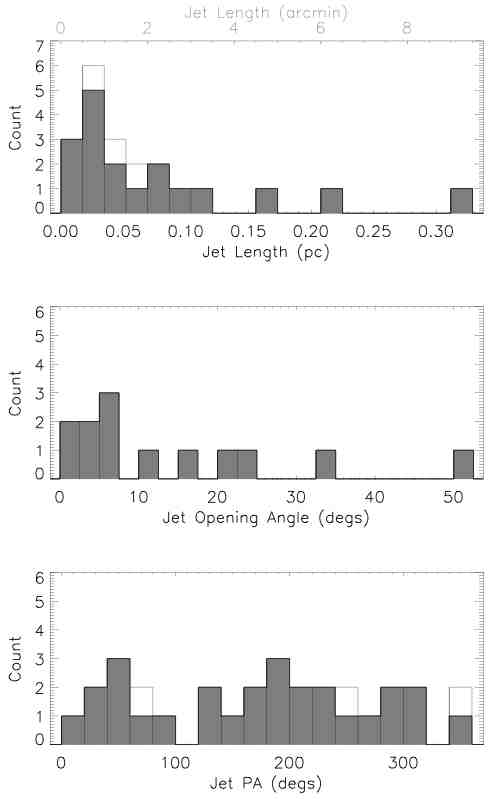}
\caption{Distribution of jet lengths (in 0.5\arcmin \ bins) (\textit{top}),  jet opening angles (in 2.5\degr \ bins) (\textit{middle}), and jet position angles (in 20\degr \ bins) (\textit{bottom}). The open columns are for all \mybf{molecular hydrogen outflows} in Ophiuchus identified by this paper and the gray filled columns are for the previously known \mybf{molecular hydrogen outflows} in Ophiuchus.}
\label{figparahist}
\end{figure}

Ophiuchus is located considerably closer to us \citep[119\,pc,][]{lombardi08} than Perseus \citep[$\sim$300\,pc,][]{perbook} and Orion A \citep[$\sim$420\,pc,][]{menten07}. Of our \mybf{21 molecular hydrogen outflows}, \mybf{12 (57\%)} exceed one arcminute in length (at a distance of 119\,pc, 1\arcmin \ is equivalent to 0.035\,pc) and 4 (\mybf{19\%}) exceed three arcminutes (corresponding to 0.1\,pc at a distance of 119\,pc). A histogram (top panel in figure \ref{figparahist}) shows the distribution of jet lengths. Shorter jets are more common than longer jets. \mybf{The median of jet lengths is 1.1\arcmin, corresponding to the physical scale of about 0.04\,pc in Ophiuchus.}
 
\mybf{\citet{davis09} identified 116 jets in Orion A and the lengths of these 116 jets show an exponential frequency distribution.} While this distribution \citep[see][Fig.~14a]{davis09} is biased towards shorter flow lengths as longer jets may become invisible in H$_{2}$ emission once they exit their molecular cloud, \citet{davis09} still suggests that the H$_{2}$ flows in Orion A are preferentially short because inclination alone would result in a frequency distribution increasing towards longer jet length if one assumes that all jets have equal length and are inclined randomly.
 \mybf{\citet{ioa1,ioa2} detected 134 MHOs in Serpens and Aquila region using the UKIRT telescope and associated them with 131 molecular hydrogen outflows. They found that the projected jet length drops exponentially in number for longer jets, and does not behave as a power law. They fitted the distribution of jet lengths with the exponential function of 10$^{-0.75\times length(pc)}$ and also found that a simple Monte Carlo-type model of jets with speeds of 40-130 km s$^{-1}$ and ages between 4 and 20 $\times$ 10$^3$ yr can reproduce the observed length distribution.}
 \mybf{The distribution of jet lengths obtained by us in Ophiuchus also shows a decaying distribution for the jets with length $>$ 0.5\arcmin. However, our sample is too small for a statistical analysis.}


Figure \ref{figparahist} shows the distribution of jet opening angles (middle panel) and the distribution of jet position angles (bottom panel). Note that the jet position angles show a nearly homogeneous distribution \mybf{(a Kolmogorov-Smirnov (KS) test shows that there is a $>$\,97\% probability that such a distribution is drawn from a homogeneously distributed sample)}, indicative of molecular hydrogen outflows in Ophiuchus being oriented randomly. \mybfnew{This is in agreement with the findings for Perseus \citep{davis08} and Orion A \citep{davis09} \mybf{but different from the results from DR21/W75 \citep{davis07}
. \citet{davis07} have found that the H$_2$ outflows, in particular from massive cores, are preferentially orthogonal to the molecular ridge. 
The alignment of position angles of outflows to some preferred direction seems to indicate a physical connection between PAs of outflows and the large-scale cloud structure \citep{davis07}. }}

\mybf{For the random orientation of outflows in Orion A, \citet{davis09} discussed its possible explanation. The orientation of outflows could be connected to the position angles of cores based on the standard paradigm that clouds collapse along field lines to form elongated, clumpy filaments from which chains of protostars are born, with the associated outflows aligned parallel with the local field, and perpendicular to the chains of cores \citep{mous76,bane06,davis09}. To test this paradigm in detail, \citet{davis09} selected a sub-sample of 22 H$_2$ jets that are associated with 1.1 mm dust cores. They surprisingly found that even from this finely-tuned sample, the distribution of H$_2$ jet position angles with respect to core PAs is completely random. They suggested that the random orientation of outflows may be caused by the relatively poor spatial resolution of the millimeter observations which can not disentangle the molecular envelopes associated with multiple sources.}

\mybf{Of our 21 jets in Ophiuchus, only 4 jets are associated with 1.1 mm dust cores. We checked these four jets and found that there is no obvious correlation between the orientation of outflows and the position angles of dust cores, which agrees with the results from \citet{davis09}.}

\begin{figure}[htb]
\centering
\includegraphics[scale=0.5]{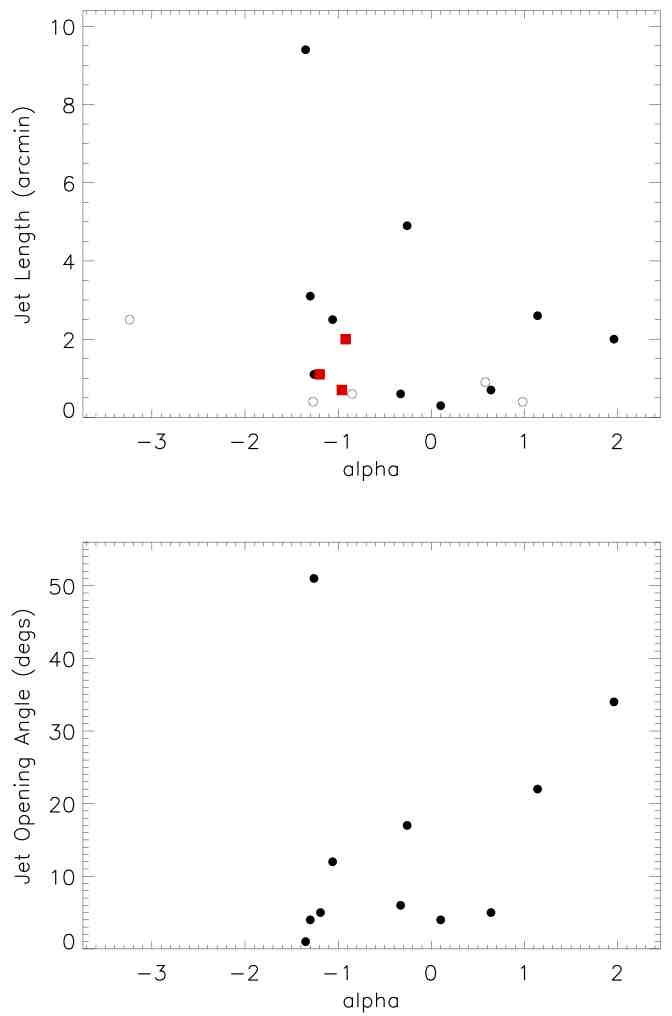}
\caption{H$_2$ jet length (\textit{top}) and jet opening angle (\textit{bottom}) plotted against source spectral index, $\alpha$. The black filled circles represent the previously known H$_2$ outflows which have both jet length and jet opening angle measurements. The black open circles represent the previously known H$_2$ outflows which have only jet length measurements. The outflows that consist of newly detected MHOs are marked with red filled squares.}
\label{figparavsalpha}
\end{figure}

Figure \ref{figparavsalpha} (top panel) shows the relation of jet lengths vs.\ spectral indices of the driving sources, which does not exhibit any significant correlation. The spectral index of the longest jet has a value of -1.35. For the Class 0 source VLA 1623-243, which drives the second longest outflow with a length of 6.2\arcmin, a spectral index from fitting the SED from 2\,$\mu$m to 24\,$\mu$m can not be obtained as this source is not detected in the IRAC observations. Of course, we cannot expect a simple linear relation between jet lengths and spectral indices of driving sources because the relation between them should be complex: for the very young sources, we should expect a short jet length as a jet should start from the length of zero; then jets probably expand, and later on get shorter again as force support fades \mybf{\citep{phdstanke}}. However, we didn't see any similar correlation between jet lengths and spectral indices of driving sources from Figure \ref{figparavsalpha}, either. The mean length of jets driven by protostars ($\alpha > -0.3$, including Class 0/I and Flat sources) is about \mybf{2.1\arcmin}, which is basically the same as the mean length of jets driven by Class II/III sources of about \mybf{2.1\arcmin}. Similar results have been found in Perseus \citep{davis08} and Orion A \citep{davis09}. The H$_{2}$ jets in Perseus and Orion A also show no correlation between jet lengths and flow source spectral indices.

Millimeter observations indicate that protostars, which are in an earlier stage of evolution, have an increased likelihood to drive more collimated CO outflows \citep{lee02,as06}. For \mybf{molecular hydrogen outflows}, however, no such trend has been observed. \citet{davis08,davis09} found no correlation between jet opening angle and the age (via the proxy of spectral index) of the driving source in Perseus and Orion A. Our sample of \mybf{molecular hydrogen outflows} in Ophiuchus also does not exhibit a correlation between jet opening angle and flow source spectral index (see Figure \ref{figparavsalpha}, bottom panel). \mybfagain{In Figure \ref{figparavsalpha}, eleven previously known H$_2$ outflows, which have both jet length and jet opening angle measurements, are marked with black filled circles. Five previously known H$_2$ outflows, which have only one H$_2$ feature for each and thus their outflow opening angles can not be calculated, are labeled with black open circles. The three outflows that consist of newly detected MHOs are marked with red filled squares. These three outflows also each have only one H$_2$ feature. Note that in the top panel there are 3 outflows with the same jet length (1\farcm1) and similar spectral indices (-1.26, -1.2, and -1.19, see table.~\ref{promho}). Note also that there are two outflows without spectral indices because their driving sources are invisible in the IRAC bands and thus their spectral indices can not be obtained by fitting the fluxes from 2\,$\mu$m to 24\,$\mu$m. For the molecular hydrogen outflows in Ophiuchus,} the mean opening angle of jets driven by protostars ($\alpha > -0.3$) is about \mybf{15.7} degs and the mean opening angle of jets driven by Class II/III sources ($\alpha < -0.3$) is about \mybf{13.6} degs.

\citet{davis09} analyzed the reasons for a lack of correlation between $\alpha$ and jet length or jet opening angle. Firstly, shock-excited H$_{2}$ emission is not a good tracer for outflow parameters because of its short cooling time, making H$_{2}$ emission not a sensitive tracer for long jets. In addition, the wings of jet-driven bow shocks are often much wider than the underlying jet and changes in flow direction due to precession.
These facts could result in the under-estimation of jet length and the over-estimation of jet opening angle. Secondly, the precise relationship between source spectral index and source age has been not established, yet.
Our sample in Ophiuchus 
is somewhat biased towards shorter jets compared to Perseus or Orion A \mybf{because our observations just cover part of the $\rho$ Ophiuchi cloud. On the one hand, our survey only covers part of area of the L1688 core, while the survey of \citet{davis08} covers the whole western portion of Perseus and the survey of \citet{davis09} covers almost the entire Orion A molecular ridge; on the other hand, the typical field of view of SofI is about 0.2$\times$0.2\,pc$^2$ at the distance of 119\,pc, while the typical field of view of WFCAM is about 6$\times$6\,pc$^2$ at the distance of 420\,pc and about 5$\times$5\,pc$^2$ at the distance of 300\,pc.} Therefore it is more unlikely for our sample to exhibit a correlation between  $\alpha$ and jet length.

\section{Discussion}
\subsection{What kind of YSOs drive H$_2$ outflows?}
\begin{figure}[htb]
\centering
\includegraphics[scale=0.5]{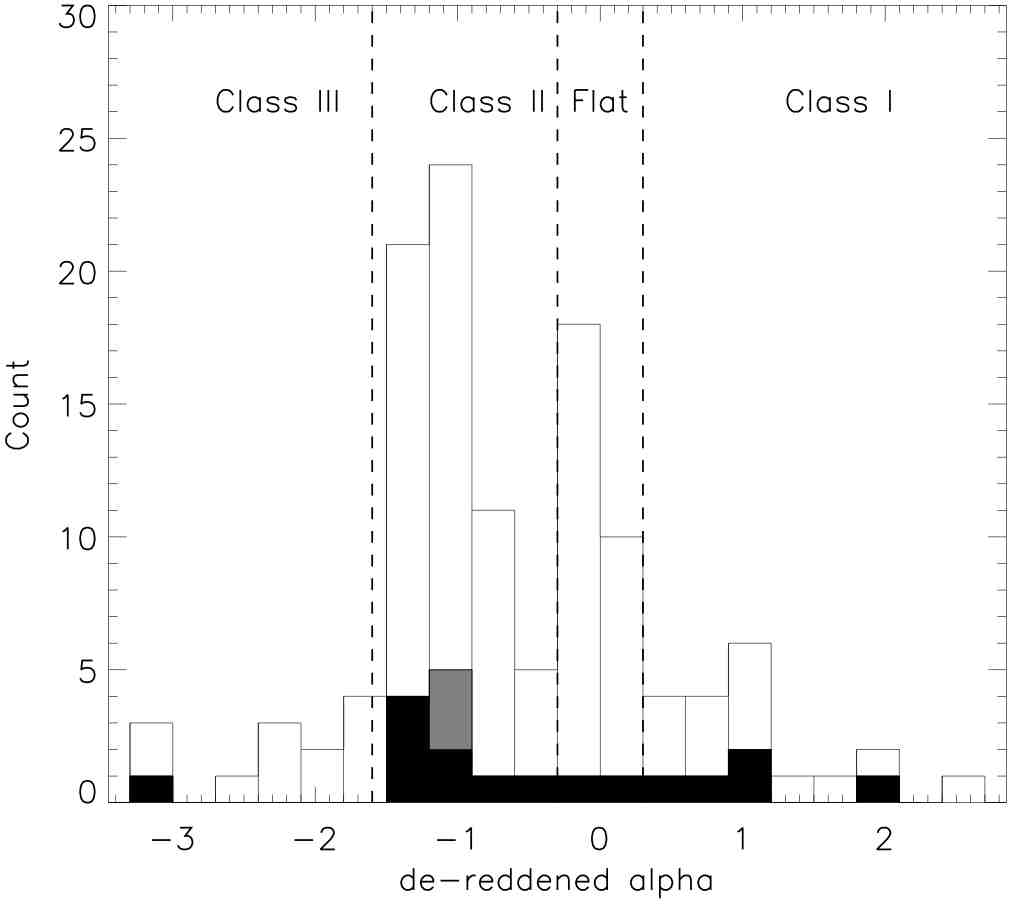}
\caption{Histogram of the de-reddened spectral indices (in bins of 0.3) of all YSOs \citep{evans09} located inside the polygon on Fig.~\ref{figstat} (open columns), all H$_{2}$ flow sources identified by this paper (filled gray columns), and H$_{2}$ flow sources excluding those associated with the three newly detected flows (filled black columns) 
 . The dashed lines show the criteria of YSO classification from \citet{greene94}.}
\label{figysohist}
\end{figure}

Of our \mybf{21} driving sources of \mybf{the molecular hydrogen outflows}, 7 \mybf{(33\%)} are Class 0/I sources; 2 \mybf{(10\%)} are flat spectrum sources; \mybf{11 (52\%)} are Class II sources and 1 \mybf{(5\%)} is a Class III source \citep[criterion for YSO classification is from ][]{greene94}. Figure \ref{figysohist} shows the distribution of spectral indices for the outflow sources and all the YSOs in the ``complete" region shown in Fig. \ref{figstat} with a solid polygon. Obviously, outflow sources and all YSOs have a similar distribution of spectral indices \mybf{(a Kolmogorov-Smirnov (KS) test shows that there is a $>$\,95\% probability that these two distributions are drawn from the same distributed sample)}, which means that in Ophiuchus the \mybf{molecular hydrogen outflows} can be driven by YSOs at any evolutionary stage. This result is different from the previous studies.

In Perseus, Orion A and other distant high mass star forming regions such as W75/DR21, the majority of \mybf{molecular hydrogen outflows} are driven by protostars with the positive spectral indices \citep{davis07,kumar07,davis08,davis09}. The mean values of spectral indices of outflow sources for Perseus, Orion A and W75/DR21 are 1.4, 0.86 and 1.9, respectively. The mean value of our flow source sample in Ophiuchus is \mybf{-0.52}, which corresponds to Class II sources. However, there are several uncertainties in determining YSO spectral indices. The YSO spectral indices in WR75/DR21 were obtained using only IRAC photometry, which would result in the over-estimation of the mean value of $\alpha$ \citep{kumar07,davis09}. Moreover, the source spectral indices used in this work are the de-reddened values from \citet{evans09}. In fact, if we use the apparent spectral indices, the mean value of $\alpha$ of flow sources in Ophiuchus will change to \mybf{-0.16}. Certainly, even though considering these uncertainties, the value of \mybf{-0.16} is still significantly lower than the mean values observed in three other star-forming regions. Actually, the biggest difference between our flow source sample in Ophiuchus and other samples in Perseus, Orion A or W75/DR21 is that more than half of the \mybf{21} driving sources of H$_{2}$ jets in Ophiuchus are Class II sources.

Results from the Spitzer \textit{c2d} project have shown that the Ophiuchus and Perseus molecular clouds have nearly the same proportion of protostars (Class I and flat-spectrum sources) to Class II sources, i.e. 40\% in Ophiuchus and 45\% in Perseus \citep[see Table 5 in][]{evans09}. Therefore, the observed difference between Ophiuchus and Perseus regarding the association of \mybf{molecular hydrogen outflows} with protostars and Class II sources is not caused by different proportions between protostars and Class II sources in these two regions. \mybf{The difference between the Ophiuchus molecular cloud, where most \mybf{molecular hydrogen outflows} are driven by Class II sources, and the Perseus, Orion A, and other distant high mass star forming regions such as W75/DR21, where the majority of \mybf{molecular hydrogen outflows} are observed to be driven by protostars, can be explained with the observational selection effects.}

\begin{figure}[htb]
\centering
\includegraphics[scale=0.5]{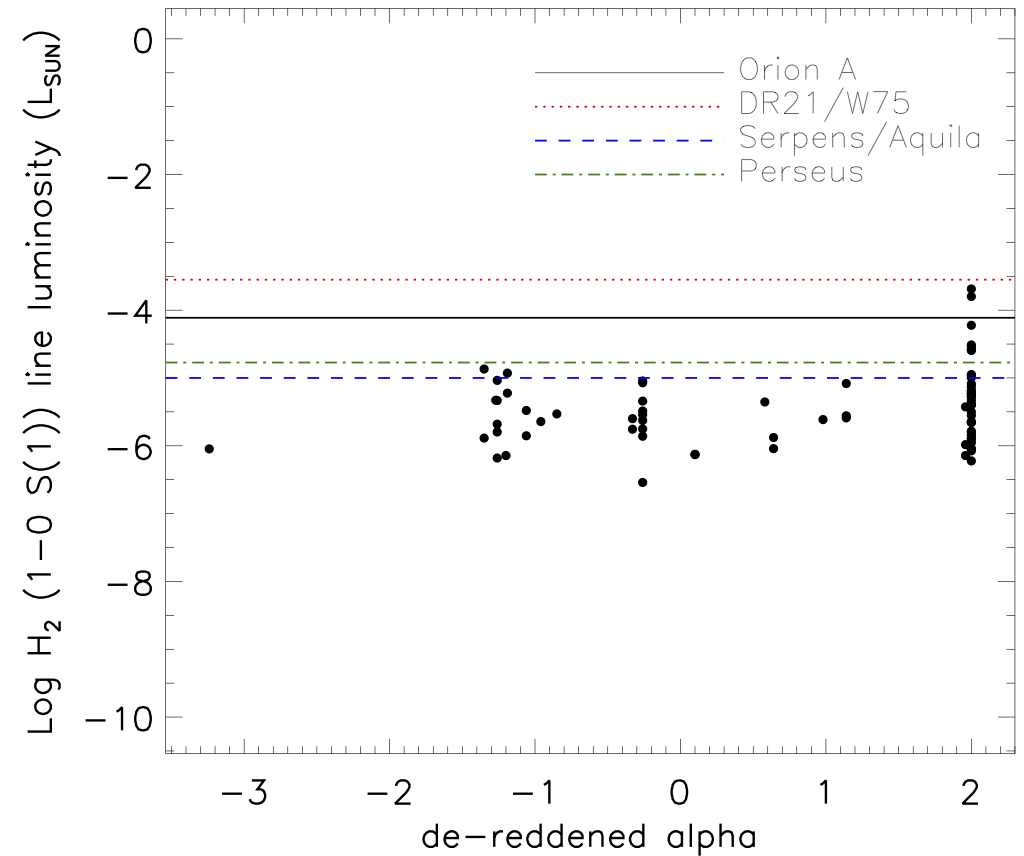}
\caption{H$_2$ 1-0 S(1) line luminosity plotted against source spectral index, $\alpha$, for the MHO features in Ophiuchus. The black solid line, the red dotted line, the blue dashed line, and the green dot-dashed line show the detection limit from \citet{davis09}, \citet{davis07}, \citet{ioa1,ioa2}, and \citet[][private communication]{davis08}, respectively.}
\label{figfluxvsalpha}
\end{figure}

\mybf{We converted the fluxes of the MHO features in Ophiuchus to the H$_2$ 1-0 S(1) line luminosity with the conversion of $L_{H_{2}} = F_{H_{2}} \times 4\pi d^2$. We adopted 119\,pc \citep{lombardi08} as the distance ($d$) of Ophiuchus. Note that we did not consider the effect of extinction. Figure.~\ref{figfluxvsalpha} shows the relation of H$_2$ 1-0 S(1) line luminosities vs.\ spectral indices of their driving sources. We found that the median of H$_2$ 1-0 S(1) line luminosity of the MHO features in Ophiuchus is about 3$\times$10$^{-6}$ L$_{\odot}$. Note that there are two YSOs ([EDJ2009] 800 and VLA 1623-243, see Table.~\ref{promho}) with no spectral indices because these two sources are invisible in IRAC bands and their spectral indices can not be obtained by fitting the flux from 2\,$\mu$m to 24\,$\mu$m. Here we give the value of 2 for the spectral indices of [EDJ2009] 800 and VLA 1623-243 just in order to plot the H$_2$ line luminosities of the MHO features driven by them on the figure. The lines with different colors and styles show the different detection limits from \citet{davis09}, \citet{davis07}, \citet{ioa1,ioa2}, and \citet{davis08}. The black solid line shows the detection limit of the H$_2$ survey towards the Orion A molecular ridge \citep{davis09}. The dot-dashed line shows the detection limit of the H$_2$ survey towards Perseus \citep[][private communication]{davis08}, while the red dotted line shows the detection limit of the H$_2$ survey towards DR21/W75 \citep{davis07}. The blue dashed line shows the detection limit of the H$_2$ survey towards the Serpens/Aquila region on the Galactic plane \citep{ioa1,ioa2}. Note that we have converted the flux detection limits to the luminosity units assuming that Orion A is located at the distance of 420\,pc and Perseus is located at the distance of 300\,pc \citep{perbook} as well as DR21/W75 and Serpens/Aquila are both located at the distance of $\sim$3\,kpc. Obviously, most of MHO features in Ophiuchus have lower H$_2$ 1-0 S(1) line luminosities than the detection limits of the surveys towards Orion A, Perseus, DR21/W75, and Serpens/Aquila. In other words, the $\rho$ Ophiuchi molecular cloud is the nearest star forming region to us and therefore, we have a larger possibility to detect the faint \mybf{molecular hydrogen outflows} from T Tauri stars. With deeper H$_{2}$ imaging we can expect to detect many faint MHOs driven by T Tauri stars in Perseus, Orion A and other distant star forming regions such as DR21.}

There are 123 YSOs in the L1688 core (just our ``complete" region which is shown in Fig.~\ref{figstat}), including 19 Class 0/I sources, 30 Flat-spectrum sources, 64 Class II sources and 10 Class III sources \citep[only the identifications from Spitzer,][]{evans09}. \mybf{This classification is based on the de-reddened spectral indices of the YSOs in Ophiuchus. In deriving these de-reddened spectral indices, \citet{evans09} assumed a value of A$_V$ = 9.76 mag for extinction toward all YSOs in Ophiuchus. So the numbers of YSOs in different evolutionary stages (Class I, Flat, II, and III) would change if we adopt another extinction assumption.}

\mybf{The value of A$_V$ = 9.76 mag is the mean extinction of the whole $\rho$ Ophiuchi molecular cloud, but our survey only covers the partial region of L1688. In fact, the L1688 core has much higher extinction than other parts of the $\rho$ Ophiuchi cloud. The extinction map taken from \textit{c2d} final delivery data\footnote{http://data.spitzer.caltech.edu/popular/c2d/20071101\_enhanced\_v1/\\
oph/extinction\_maps/OPH\_270asec\_Av.fits} shows that the L1688 core region has visual extinction ranging from 10 mag to 35 mag. For L1688 core, \citet{evans09} recalculated the de-reddened spectral indices for Class I and Flat sources using different extinction corrections and they found }
\mybf{that about 36\% flat-spectrum sources in Ophiuchus would de-redden to Class II sources if a more realistic extinction correction is applied to the estimate of the de-reddened spectral index. Using this percentage for the correction of the number of flat-spectrum sources, we estimate that about 11 of 30 flat-spectrum sources are highly reddened Class II sources. After this correction, there are 19 Class 0/I sources, 19 Flat-spectrum sources and 75 Class II sources in the L1688 core region.}

\mybf{Of course, the discussion above assumes that the same fraction of flat-spectrum sources which drives jets as the flat-spectrum sources which does not drive jets deredden to Class II sources. However, jets may highlight the ``true" flat-spectrum
sources in fact, and only the flat-spectrum sources which do not drive jets might deredden to Class II sources. Considering this, if we assume that 2 flat-spectrum sources which drive jets are ``true" flat-spectrum sources and about 10 of 28 flat-spectrum sources are highly reddened Class II sources, then there are 19 Class 0/I sources, 20 flat-spectrum sources and 74 Class II sources in the L1688 core region.}

\mybf{Of our 21 outflows, 7 are driven by Class 0/I sources, 2 are driven by ``flat spectrum" sources and 11 are driven by Class II sources. Now we can deduce that the percentage for protostars (Class 0/I$+$Flat) and T Tauri stars (Class II) to drive \mybf{molecular hydrogen outflows} is $\sim$23\% and $\sim$15\%, respectively. We can see that there is no significant difference between the percentages of protostars and T Tauri stars which drive \mybf{the molecular hydrogen outflows}. This might mean that the occurrence of \mybf{molecular hydrogen outflows} is not correlated with the age of the driving sources.}

\mybf{Here we should mention the works of \citet{hat07}. \citet{hat07} investigated the CO molecular outflows in $^{12}$CO (J = 3-2) towards 51 submilimetre cores in the Perseus molecular cloud using James Clerk Maxwell Telescope (JCMT). They calculated the outflow momentum fluxes for all the possible outflow sources and they found that outflow power may not show a simple decline between the Class 0 to Class I stages. These results might indicate that the evolution of outflows is not a simple process.}

\subsection{Velocity variations along the outflows}
Proper motion studies for HH 34 have found a systematic decrease in proper motions with distance from the driving source \citep{devine97}. To explain this velocity decrease, \citet{cr00} modeled HH 34 and found that the variation in velocity in HH 34 is most likely to be the result of environmental drag on the propagation of individual jet knots.

\begin{figure}[htb]
\centering
\includegraphics[scale=0.6]{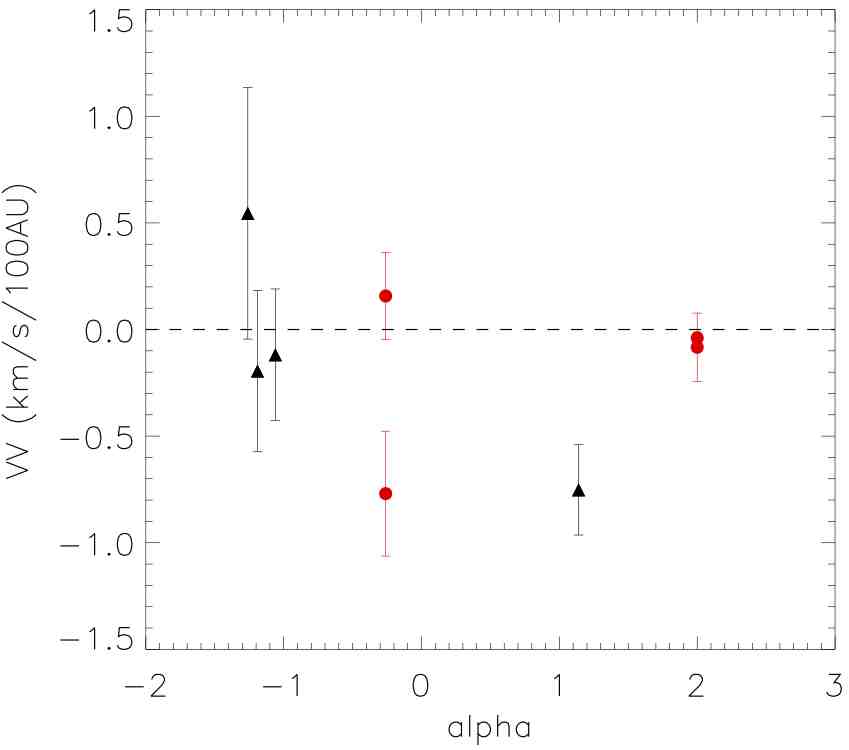}
\caption{Variation in velocity along outflow plotted against de-reddened spectral index of outflow source. \mybf{The VVs of each lobe for the bipolar outflows are marked with the red circles while the VVs for the single-lobe outflows are labeled with the black triangles. The error bars represent the uncertainties of the linear fittings.}} 
\label{figvv}
\end{figure}

We also investigated the possible velocity variation along outflows in our MHO sample. \mybf{Out of 21 molecular hydrogen outflows, 5 single-lobe outflows have at least two features with the reliable PM measurements while 2 bipolar outflows have at least two features with the reliable PM measurements in any one of two lobes. For each lobe of these 7 outflows, we calculated the standard deviation of the velocities. We found that the median value of the standard deviations is about 8\,km s$^{-1}$ and the minimum and maximum values are 4 and 47 km s$^{-1}$, respectively.} In order to describe the variation in velocity along outflows (VV), we fitted the distribution of velocities of MHO features with the distances from the driving sources using a linear least square method. \mybf{For the bipolar outflows such as the outflow driven by VLA 1623-243 and the outflow driven by YLW 52, we calculated the VVs for each lobe separately.} \mybf{The negative value of VV means proper motions of MHO features decreasing with the distances from the driving sources.} Fig.~\ref{figvv} shows the variation in velocity along outflow plotted against the de-reddened spectral index of outflow sources. \mybf{The VVs of each lobe for the bipolar outflows are marked with the red circles while the VVs for the single-lobe outflows are labeled with the black triangles. The error bars represent the uncertainties of the linear fittings. We also exclude a value of VV with very large error bar (the outflow driven by [GY92] 93). Note that for the bipolar outflow which consists of MHO 2105 and MHO 2106, its \mybfnew{driving source} is a Class 0 source (VLA 1623-243) whose spectral index can not be obtained due to the lack of IRAC detection. Here we give a value of 2 for the spectral index of VLA 1623-243 with the aim to plot the VVs of the outflow driven by VLA 1623-243 in the figure.} 

Given the fitting error, \mybf{two} H$_{2}$ jets show significant decrease in proper motions along the outflow. Most of \mybf{molecular hydrogen outflows} in Ophiuchus have roughly constant proper motions with increasing distance from the outflow sources. This result agrees with the finding by \citet{mcg07}. \citet{mcg07} investigated the proper motions of several HH objects from classic T Tauri stars (CTTS) and they found that the velocity appears to be constant despite the large distances between consecutive HH objects. The maximum dynamic age of the HH objects investigated by \citet{mcg07} is about $\sim$10$^3$ years. \mybf{\citet{mcg07} suggested two possibilities for this result: one is that the ejection velocity at the source was much higher $\sim$10$^3$ years ago when the more distant objects were ejected and has decreased over the intervening years to its current velocity. Meanwhile the velocity of the older/more distant objects has decreased over time via interactions with the parent cloud; the other is that the velocity at the source has remained at approximately the same value over the last 10$^3$ years and the distant objects have not been slowed down much by their interaction with the ambient medium.}
\mybf{The maximum dynamic age of our \mybf{molecular hydrogen outflows} is about 10$^4$ years. Based on our results, we also can not confirm or exclude any one of these two possibilities.}



\subsection{Future work}
The $\rho$ Ophiuchi molecular cloud is the nearest star forming region. It is an ideal laboratory for the studies of star formation. We identified \mybf{21} driving sources for \mybf{27} MHOs in Ophiuchus, which offered a good sample for future high-resolution observations. According to jet launching models, the jet launching zones of MHOs are within 0.5\,AU from the driving sources, which correspond to around 40\,mas at the distance of Ophiuchus. This requirement for angular resolution can be achieved, e.g., with ground-based interferometric instruments like GRAVITY, which is an adaptive optics assisted, near-infrared VLTI instrument for precision narrow-angle astrometry and interferometric phase referenced imaging of faint objects \citep{gravity}.

With an accuracy of 10 micro-arcsecond for astrometry and 4 mas for resolution, GRAVITY/VLTI could study in detail the environment within 0.5\,AU from young stars. The largest proper motion of MHO features which we detected in Ophiuchus is about \mybf{247} mas/yr ,which corresponds to about \mybf{28} $\mu$as per hour. Therefore for some MHO features with high proper motions we can expect to study the jets ``in real time" with the resolution of GRAVITY/VLTI. The \mybf{molecular hydrogen outflows} discussed in the present paper could serve as targets for high precision astrometric monitoring in order to study the launching zone of MHOs, the collimation mechanism for central jets, and the physics of shocks.

\section{Summary}
We performed a deep near-infrared search for \mybf{molecular hydrogen emission line objects} (MHOs) towards the $\rho$ Ophiuchi molecular cloud, covering an area of $\sim$ 0.11 deg$^2$. In total, we discovered \mybf{six} new MHOs and detected \mybf{32} known MHOs. Using previously-published H$_{2}$ images, we measured the proper motions for the H$_{2}$ emission features in \mybf{32 MHOs}. The proper motions lie in the range of \mybf{14 to 247\,mas/yr}, corresponding to the transversal velocities of \mybf{8 to 140\,km/s} with the \mybf{median} tangential velocity of about \mybf{35\,km/s}.

Based on outflow morphology and PM analysis, we associated \mybf{27 MHOs} with \mybf{21} driving sources. Out of the \mybf{21} outflow sources, \mybf{two} also drive CO outflows, \mybf{four (19\%)} have 850 $\mu$m and four \mybf{(19\%)} have 1.1\,mm dust core counterparts. The \mybf{molecular hydrogen outflows} have the \mybf{median} length of \mybf{$\sim$0.04\,pc} and are oriented randomly. H$_{2}$ jet lengths (\textit{L}) and H$_{2}$ jet opening angles (\textit{$\theta$}) show no correlations with the spectral indices $\alpha$ of the driving sources, which serve as a proxy for the ages of the outflow sources.

Among the \mybf{21 molecular hydrogen} outflows identified in Ophiuchus, \mybf{11} outflows emanate from Class II sources. Taking the high extinction in our H$_{2}$ emission survey region into account, we estimate that 23\% and \mybf{15\%} of the \mybf{protostars} and \mybf{T Tauri stars}, respectively, drive \mybf{molecular hydrogen outflows}. The comparable percentages for protostars and T Tauri stars in Ophiuchus which drive H$_{2}$ outflows are in contrast with previous results in the Perseus, Orion A, and W75/DR21 clouds, where most of the H$_{2}$ outflows are driven by protostars. The difference can be explained with an observational selection effect, as \mybf{most of the MHO features detected by us in Ophiuchus have  low H$_2$ 1-0 S(1) line luminosities which are under the detection limit of the surveys towards Orion A, Perseus, DR21/W75, and the Serpens/Aquila region on the Galactic plane.} 


In most cases the proper motions of \mybf{MHO features} in \mybf{molecular hydrogen outflows} do not vary with distance from the outflow source\mybf{, which agrees with the previous studies such as \citet{mcg07}.} 

\begin{acknowledgements}
We thank S. Zhang for the IDL plotting code which has been uploaded to the Website\footnote{http://code.google.com/p/aicer/}. We also thank C. J. Davis for providing the MHO nomenclature for the new H$_{2}$ knots. H. Wang acknowledges the support by NSFC through grants 11173060, 10921063, and 10733030. We have made extensive use of the AstrOmatic (http://www.astromatic.net/software) analysis tools SEXTRACTOR\citep{sex} and SWARP\citep{swarp} . We also use the Starlink analysis tool TOPCAT (http://starlink.jach.hawaii.edu/) and the WCSTools package (http://tdc-www.harvard.edu/software/wcstools/index.html). This research has also made use of the SIMBAD database, operated at CDS, Strasbourg, France. In particular, the Aladin image server is very useful in locating previously known objects on our frames.
\end{acknowledgements}


\begin{landscape}
\begin{table}[htb]

\centering
\caption[]{\centering Driving sources and properties of molecular hydrogen outflows in Ophiuchus}
         \label{promho}
\begin{tabular}{lccccccccccclc}
\hline\hline
Outflow source& [EDJ2009]\tablefootmark{a}&RA\tablefootmark{a} & Dec.\tablefootmark{a} & $\alpha$\tablefootmark{a} & MHO\tablefootmark{b} & 850 $\mu$m\tablefootmark{c} & 1100 $\mu$m\tablefootmark{d} & \textit{L}\tablefootmark{e}  & \textit{$\theta$}\tablefootmark{f} & PA\tablefootmark{g} & DA\tablefootmark{h}  & VV\tablefootmark{i} & CO outflow\tablefootmark{j}\\
    &   &   (J2000) & (J2000)&  & & core? & core? & (\arcmin) & (deg) & (deg) & (10$^3$ yr) & (km s$^{-1}$ 100AU$^{-1}$) & associated?\\ \hline
[EDJ2009]800 &800 & +16 26 14.6 & -24 25 07.5 &\ldots  &2102 &Y &N &0.9 &1 &141 &2.9 &\ldots &N \\                                                       
YLW31 &806 & +16 26 18.9 & -24 28 19.6 &-1.30 &2103-2104 &Y &N &3.1 &4 &62 &\ldots &\ldots &N \\                                                              
CRBR2317.5-1729&807 & +16 26 19.0 & -24 24 14.2 &-0.33 &2150 &N &N &0.6 &6 &48 &0.9 &\ldots &N \\                                                    
GSS32 &820 & +16 26 24.0 & -24 24 48.0 &-0.96 &2151 &N &N &0.7 &\ldots&252 &2.0 &\ldots &N \\{}                                                          
[GY92]30 &822 & +16 26 25.5 & -24 23 01.3 &0.64 &2137 &N &Y &0.7 &5 &120 &0.5 &\ldots &N \\                                                           
VLA1623-243 &825 & +16 26 26.4 & -24 24 30.0 &\ldots &2105-2106 &Y &N &6.2 &23 &123 &2.7 &-0.1$\pm$0.2;-0.04$\pm$0.03 &Y \\{}                
[GY92]93 &833 & +16 26 41.2 & -24 40 17.9 &-1.35 &2115a,c &N &N &9.4 &1 &40 &9.8 &-46$\pm$131 &N \\                                   
WL18 &841 & +16 26 49.0 & -24 38 25.1 &-1.27 &2108 &N &N &0.4 &\ldots&297 &\ldots &\ldots &N \\                                                              
YLW5 &857 & +16 27 02.3 & -24 37 27.2 &0.58 &2109 &N &Y &0.9 &\ldots&297 &1.9 &\ldots &N \\                                                               
BBRCG24 &869 & +16 27 09.1 & -24 34 08.0 &-1.26 &2115d,e,b,f,g &N &N &1.1 &51 &167 &0.7  &0.5$\pm$0.6 &N \\{}       
[GY92]232 &875 & +16 27 13.3 & -24 41 33.4 &-3.24 &2114 &N &N &2.5 &\ldots&163 &3.6 &\ldots &N \\{}                                                       
[GY92]235 &877 & +16 27 13.8 & -24 43 31.6 &-0.85 &2113 &N &N &0.6 &\ldots&225 &1.3 &\ldots &N \\{}                                                       
[GY92]239 &880 & +16 27 15.4 & -24 26 39.6 &-1.20 &2153 &N &N &1.1 &\ldots&77 &1.1 &\ldots &N \\                                                         
YLW14 &891 & +16 27 21.8 & -24 29 53.1 &0.98 &2116 &N &N &0.4 &\ldots&40 &0.3 &\ldots &N \\                                                               
WSB49 &893 & +16 27 23.0 & -24 48 07.1 &-1.19 &2111 &N &N &1.1 &5 &183 &1.8 &-0.2$\pm$0.4  &N \\                                                
YLW16 &901 & +16 27 28.0 & -24 39 33.4 &1.96 &2118-2119 &Y &Y &2.0 &34 &18 &0.6 &\ldots &Y \\                                                      
YLW47 &916 & +16 27 38.3 & -24 36 58.4 &-1.06 &2122,2123c &N &N &2.5 &12 &195 &6.2 &-0.1$\pm$0.3  &N \\{}                                    
[GY92]344 &932 & +16 27 45.8 & -24 44 53.8 &0.10 &2126 &N &N &0.3 &4 &20 &0.4 &\ldots &N \\                                                           
YLW52 &937 & +16 27 51.8 & -24 31 45.4 &-0.26 &2128-2131 &N &N &4.9 &17 &81 &8.8 &0.2$\pm$0.2;-0.8$\pm$0.3 &N \\         
YLW58 &950 & +16 28 16.5 & -24 36 57.9 &-0.92 &2154 &N &N &2.0 &\ldots&357 &\ldots &\ldots &N \\                                                            
MMS126 &954 & +16 28 21.6 & -24 36 23.4 &1.14 &2132 &N &Y &2.6 &22 &37 &1.2  &-0.8$\pm$0.2 &N \\                                                
\hline
\end{tabular}
\tablefoot{
\tablefoottext{a}{Name, position and de-reddened spectral index of driving sources from \citet{evans09}.}
\tablefoottext{b}{The MHOs, which are driven by the driving sources. MHO names are from \citet{davis10}.}
\tablefoottext{c}{``Y" or ``N'' indicates whether or not a 850 $\mu$m \ dust core is associated with the driving source. The details on the 850 $\mu$m \ dust cores can be found in  \citet{johnstone2000}.}
\tablefoottext{d}{Same as for the 850 $\mu$m \ cores, but for the 1100 $\mu$m \ dust cores. Details can be found in \citet{evans09} and \citet{young06}.}
\tablefoottext{e}{The entire length (L) of the outflow. This length is the distance from the emission feature to the source projected to the bisector of the opening angle in cases where more than one feature is identified in the outflow. In case there is only one feature identified in the outflow, this length is just the distance from the source to the feature.}
\tablefoottext{f}{The opening angle ($\theta$) of the outflow. This angle is measured from a cone with the smallest vertex angle centered on the driving source that includes all MHO features in the outflow. Note that the opening angle can not be calculated if there is only one feature in the outflow.}
\tablefoottext{g}{The position angle (PA) of the outflow measured east of north as computed from the bisector of the opening angle. Note that for bipolar outflows we select the angle smaller than 180 degrees as the position angle.}
\tablefoottext{h}{The dynamic age (DA) of the outflow, which is the maximum dynamic age of the features with proper motions.}
\tablefoottext{i}{Variation in velocity (VV) along the outflow. This parameter is calculated through fitting the tangential velocity of MHO features vs.\ the distance from MHO features to the driving source using a linear least square method. VV equals 1 means that the MHO features will speed up by \mybf{1 km/s} after they propagate \mybf{100\,AU} along the outflow. \mybf{Note that for the bipolar outflows, we calculated the VVs for each lobe.}}
\tablefoottext{j}{``Y" or ``N'' indicates whether or not a CO molecular outflow is associated with the MHO. The CO molecular outflows are identified by \citet{naka11} based on the CO (J = 3-2) and CO (J = 1-0) mapping observations.}
}
\end{table}
\end{landscape}

\clearpage
\bibliographystyle{aa}
\bibliography{ophref}

\begin{thebibliography}{101}
\expandafter\ifx\csname natexlab\endcsname\relax\def\natexlab#1{#1}\fi

\bibitem[{{Allen} {et~al.}(2002){Allen}, {Myers}, {Di Francesco}, {Mathieu},
  {Chen}, \& {Young}}]{allen02}
{Allen}, L.~E., {Myers}, P.~C., {Di Francesco}, J., {et~al.} 2002, \apj, 566,
  993

\bibitem[{{Alves de Oliveira} \& {Casali}(2008)}]{alv08}
{Alves de Oliveira}, C. \& {Casali}, M. 2008, \aap, 485, 155

\bibitem[{{Andre} {et~al.}(1990){Andre}, {Martin-Pintado}, {Despois}, \&
  {Montmerle}}]{and90}
{Andre}, P., {Martin-Pintado}, J., {Despois}, D., \& {Montmerle}, T. 1990,
  \aap, 236, 180

\bibitem[{{Andre} {et~al.}(1993){Andre}, {Ward-Thompson}, \&
  {Barsony}}]{andre93}
{Andre}, P., {Ward-Thompson}, D., \& {Barsony}, M. 1993, \apj, 406, 122

\bibitem[{{Andrews} \& {Williams}(2007)}]{andr07}
{Andrews}, S.~M. \& {Williams}, J.~P. 2007, \apj, 671, 1800

\bibitem[{{Andrews} {et~al.}(2010){Andrews}, {Wilner}, {Hughes}, {Qi}, \&
  {Dullemond}}]{andr10}
{Andrews}, S.~M., {Wilner}, D.~J., {Hughes}, A.~M., {Qi}, C., \& {Dullemond},
  C.~P. 2010, \apj, 723, 1241

\bibitem[{{Arce} \& {Sargent}(2006)}]{as06}
{Arce}, H.~G. \& {Sargent}, A.~I. 2006, \apj, 646, 1070

\bibitem[{{Arce} {et~al.}(2007){Arce}, {Shepherd}, {Gueth}, {Lee}, {Bachiller},
  {Rosen}, \& {Beuther}}]{arce07}
{Arce}, H.~G., {Shepherd}, D., {Gueth}, F., {et~al.} 2007, Protostars and
  Planets V, 245

\bibitem[{{Bachiller}(1996)}]{bach96}
{Bachiller}, R. 1996, \araa, 34, 111

\bibitem[{{Bally} {et~al.}(2007){Bally}, {Reipurth}, \& {Davis}}]{bally07}
{Bally}, J., {Reipurth}, B., \& {Davis}, C.~J. 2007, Protostars and Planets V,
  215

\bibitem[{{Bally} {et~al.}(2008){Bally}, {Walawender}, {Johnstone}, {Kirk}, \&
  {Goodman}}]{perbook}
{Bally}, J., {Walawender}, J., {Johnstone}, D., {Kirk}, H., \& {Goodman}, A.
  2008, {The Perseus Cloud}, ed. B.~{Reipurth}, 308

\bibitem[{{Banerjee} \& {Pudritz}(2006)}]{bane06}
{Banerjee}, R. \& {Pudritz}, R.~E. 2006, \apj, 641, 949

\bibitem[{{Barsony} {et~al.}(1989){Barsony}, {Carlstrom}, {Burton}, {Russell},
  \& {Garden}}]{bbrcg}
{Barsony}, M., {Carlstrom}, J.~E., {Burton}, M.~G., {Russell}, A.~P.~G., \&
  {Garden}, R. 1989, \apjl, 346, L93

\bibitem[{{Barsony} {et~al.}(2003){Barsony}, {Koresko}, \& {Matthews}}]{bar03}
{Barsony}, M., {Koresko}, C., \& {Matthews}, K. 2003, \apj, 591, 1064

\bibitem[{{Bertin} \& {Arnouts}(1996)}]{sex}
{Bertin}, E. \& {Arnouts}, S. 1996, \aaps, 117, 393

\bibitem[{{Bertin} {et~al.}(2002){Bertin}, {Mellier}, {Radovich}, {Missonnier},
  {Didelon}, \& {Morin}}]{swarp}
{Bertin}, E., {Mellier}, Y., {Radovich}, M., {et~al.} 2002, in Astronomical
  Society of the Pacific Conference Series, Vol. 281, Astronomical Data
  Analysis Software and Systems XI, ed. {D.~A.~Bohlender, D.~Durand, \&
  T.~H.~Handley}, 228--+

\bibitem[{{Bontemps} {et~al.}(1996){Bontemps}, {Andre}, {Terebey}, \&
  {Cabrit}}]{bon96}
{Bontemps}, S., {Andre}, P., {Terebey}, S., \& {Cabrit}, S. 1996, \aap, 311,
  858

\bibitem[{{Bussmann} {et~al.}(2007){Bussmann}, {Wong}, {Hedden}, {Kulesa}, \&
  {Walker}}]{bus07}
{Bussmann}, R.~S., {Wong}, T.~W., {Hedden}, A.~S., {Kulesa}, C.~A., \&
  {Walker}, C.~K. 2007, \apjl, 657, L33

\bibitem[{{Cabrit} \& {Raga}(2000)}]{cr00}
{Cabrit}, S. \& {Raga}, A. 2000, \aap, 354, 667

\bibitem[{{Caratti o Garatti} {et~al.}(2006){Caratti o Garatti}, {Giannini},
  {Nisini}, \& {Lorenzetti}}]{cog06}
{Caratti o Garatti}, A., {Giannini}, T., {Nisini}, B., \& {Lorenzetti}, D.
  2006, \aap, 449, 1077

\bibitem[{{Casanova} {et~al.}(1995){Casanova}, {Montmerle}, {Feigelson}, \&
  {Andre}}]{cas95}
{Casanova}, S., {Montmerle}, T., {Feigelson}, E.~D., \& {Andre}, P. 1995, \apj,
  439, 752

\bibitem[{{Costa} {et~al.}(2000){Costa}, {Jessop}, {Yun}, {Santos},
  {Ward-Thompson}, \& {Casali}}]{cos00}
{Costa}, A., {Jessop}, N.~E., {Yun}, J.~L., {et~al.} 2000, in IAU Symposium,
  Vol. 200, IAU Symposium, 48P--+

\bibitem[{{Davis} \& {Eisloeffel}(1995)}]{davis95}
{Davis}, C.~J. \& {Eisloeffel}, J. 1995, \aap, 300, 851

\bibitem[{{Davis} {et~al.}(2009){Davis}, {Froebrich}, {Stanke}, {Megeath},
  {Kumar}, {Adamson}, {Eisl{\"o}ffel}, {Gredel}, {Khanzadyan}, {Lucas},
  {Smith}, \& {Varricatt}}]{davis09}
{Davis}, C.~J., {Froebrich}, D., {Stanke}, T., {et~al.} 2009, \aap, 496, 153

\bibitem[{{Davis} {et~al.}(2010){Davis}, {Gell}, {Khanzadyan}, {Smith}, \&
  {Jenness}}]{davis10}
{Davis}, C.~J., {Gell}, R., {Khanzadyan}, T., {Smith}, M.~D., \& {Jenness}, T.
  2010, \aap, 511, A24+

\bibitem[{{Davis} {et~al.}(2007){Davis}, {Kumar}, {Sandell}, {Froebrich},
  {Smith}, \& {Currie}}]{davis07}
{Davis}, C.~J., {Kumar}, M.~S.~N., {Sandell}, G., {et~al.} 2007, \mnras, 374,
  29

\bibitem[{{Davis} {et~al.}(2008){Davis}, {Scholz}, {Lucas}, {Smith}, \&
  {Adamson}}]{davis08}
{Davis}, C.~J., {Scholz}, P., {Lucas}, P., {Smith}, M.~D., \& {Adamson}, A.
  2008, \mnras, 387, 954

\bibitem[{{Dent} {et~al.}(1995){Dent}, {Matthews}, \& {Walther}}]{dent95}
{Dent}, W.~R.~F., {Matthews}, H.~E., \& {Walther}, D.~M. 1995, \mnras, 277, 193

\bibitem[{{Devine} {et~al.}(1997){Devine}, {Bally}, {Reipurth}, \&
  {Heathcote}}]{devine97}
{Devine}, D., {Bally}, J., {Reipurth}, B., \& {Heathcote}, S. 1997, \aj, 114,
  2095

\bibitem[{{Eisenhauer} {et~al.}(2008){Eisenhauer}, {Perrin}, {Brandner},
  {Straubmeier}, {Richichi}, {Gillessen}, {Berger}, {Hippler}, {Eckart},
  {Sch{\"o}ller}, {Rabien}, {Cassaing}, {Lenzen}, {Thiel}, {Cl{\'e}net},
  {Ramos}, {Kellner}, {F{\'e}dou}, {Baumeister}, {Hofmann}, {Gendron}, {Boehm},
  {Bartko}, {Haubois}, {Klein}, {Dodds-Eden}, {Houairi}, {Hormuth},
  {Gr{\"a}ter}, {Jocou}, {Naranjo}, {Genzel}, {Kervella}, {Henning}, {Hamaus},
  {Lacour}, {Neumann}, {Haug}, {Malbet}, {Laun}, {Kolmeder}, {Paumard},
  {Rohloff}, {Pfuhl}, {Perraut}, {Ziegleder}, {Rouan}, \& {Rousset}}]{gravity}
{Eisenhauer}, F., {Perrin}, G., {Brandner}, W., {et~al.} 2008, in Society of
  Photo-Optical Instrumentation Engineers (SPIE) Conference Series, Vol. 7013,
  Society of Photo-Optical Instrumentation Engineers (SPIE) Conference Series

\bibitem[{{Evans} {et~al.}(2009){Evans}, {Dunham}, {J{\o}rgensen}, {Enoch},
  {Mer{\'{\i}}n}, {van Dishoeck}, {Alcal{\'a}}, {Myers}, {Stapelfeldt},
  {Huard}, {Allen}, {Harvey}, {van Kempen}, {Blake}, {Koerner}, {Mundy},
  {Padgett}, \& {Sargent}}]{evans09}
{Evans}, N.~J., {Dunham}, M.~M., {J{\o}rgensen}, J.~K., {et~al.} 2009, \apjs,
  181, 321

\bibitem[{{Froebrich} {et~al.}(2011){Froebrich}, {Davis}, {Ioannidis},
  {Gledhill}, {Takami}, {Chrysostomou}, {Drew}, {Eisl{\"o}ffel}, {Gosling},
  {Gredel}, {Hatchell}, {Hodapp}, {Kumar}, {Lucas}, {Matthews}, {Rawlings},
  {Smith}, {Stecklum}, {Varricatt}, {Lee}, {Teixeira}, {Aspin}, {Khanzadyan},
  {Karr}, {Kim}, {Koo}, {Lee}, {Lee}, {Magakian}, {Movsessian}, {Nikogossian},
  {Pyo}, \& {Stanke}}]{uwish2}
{Froebrich}, D., {Davis}, C.~J., {Ioannidis}, G., {et~al.} 2011, \mnras, 413,
  480

\bibitem[{{Gennaro} {et~al.}(2012){Gennaro}, {Bik}, {Brandner}, {Stolte},
  {Rochau}, {Beuther}, {Gouliermis}, {Tackenberg}, {Kudryavtseva}, {Hussmann},
  {Schuller}, \& {Henning}}]{mario12}
{Gennaro}, M., {Bik}, A., {Brandner}, W., {et~al.} 2012, \aap, 542, A74

\bibitem[{{G{\'o}mez} {et~al.}(2003){G{\'o}mez}, {Stark}, {Whitney}, \&
  {Churchwell}}]{gome03}
{G{\'o}mez}, M., {Stark}, D.~P., {Whitney}, B.~A., \& {Churchwell}, E. 2003,
  \aj, 126, 863

\bibitem[{{G{\'o}mez} {et~al.}(1998){G{\'o}mez}, {Whitney}, \& {Wood}}]{gome98}
{G{\'o}mez}, M., {Whitney}, B.~A., \& {Wood}, K. 1998, \aj, 115, 2018

\bibitem[{{Grasdalen} {et~al.}(1973){Grasdalen}, {Strom}, \& {Strom}}]{gss}
{Grasdalen}, G.~L., {Strom}, K.~M., \& {Strom}, S.~E. 1973, \apjl, 184, L53+

\bibitem[{{Greene} {et~al.}(1994){Greene}, {Wilking}, {Andre}, {Young}, \&
  {Lada}}]{greene94}
{Greene}, T.~P., {Wilking}, B.~A., {Andre}, P., {Young}, E.~T., \& {Lada},
  C.~J. 1994, \apj, 434, 614

\bibitem[{{Greene} \& {Young}(1992{\natexlab{a}})}]{gre92}
{Greene}, T.~P. \& {Young}, E.~T. 1992{\natexlab{a}}, \apj, 395, 516

\bibitem[{{Greene} \& {Young}(1992{\natexlab{b}})}]{gy92}
{Greene}, T.~P. \& {Young}, E.~T. 1992{\natexlab{b}}, \apj, 395, 516

\bibitem[{{Grosso} {et~al.}(2001){Grosso}, {Alves}, {Neuh{\"a}user}, \&
  {Montmerle}}]{grosso01}
{Grosso}, N., {Alves}, J., {Neuh{\"a}user}, R., \& {Montmerle}, T. 2001, \aap,
  380, L1

\bibitem[{{Grosso} {et~al.}(2000){Grosso}, {Montmerle}, {Bontemps},
  {Andr{\'e}}, \& {Feigelson}}]{gro00}
{Grosso}, N., {Montmerle}, T., {Bontemps}, S., {Andr{\'e}}, P., \& {Feigelson},
  E.~D. 2000, \aap, 359, 113

\bibitem[{{Hatchell} {et~al.}(2007){Hatchell}, {Fuller}, \& {Richer}}]{hat07}
{Hatchell}, J., {Fuller}, G.~A., \& {Richer}, J.~S. 2007, \aap, 472, 187

\bibitem[{{Imanishi} {et~al.}(2002){Imanishi}, {Tsujimoto}, \&
  {Koyama}}]{ima02}
{Imanishi}, K., {Tsujimoto}, M., \& {Koyama}, K. 2002, \apj, 572, 300

\bibitem[{Ioannidis \& Froebrich(2012{\natexlab{a}})}]{ioa1}
Ioannidis, G. \& Froebrich, D. 2012{\natexlab{a}}, Monthly Notices of the Royal
  Astronomical Society, 421, 3257

\bibitem[{Ioannidis \& Froebrich(2012{\natexlab{b}})}]{ioa2}
Ioannidis, G. \& Froebrich, D. 2012{\natexlab{b}}, Monthly Notices of the Royal
  Astronomical Society, 425, 1380

\bibitem[{{Johnstone} {et~al.}(2000){Johnstone}, {Wilson}, {Moriarty-Schieven},
  {Joncas}, {Smith}, {Gregersen}, \& {Fich}}]{johnstone2000}
{Johnstone}, D., {Wilson}, C.~D., {Moriarty-Schieven}, G., {et~al.} 2000, \apj,
  545, 327

\bibitem[{{Kamata} {et~al.}(1997){Kamata}, {Koyama}, {Tsuboi}, \&
  {Yamauchi}}]{kam97}
{Kamata}, Y., {Koyama}, K., {Tsuboi}, Y., \& {Yamauchi}, S. 1997, \pasj, 49,
  461

\bibitem[{{Kamazaki} {et~al.}(2003){Kamazaki}, {Saito}, {Hirano}, {Umemoto}, \&
  {Kawabe}}]{kamazaki03}
{Kamazaki}, T., {Saito}, M., {Hirano}, N., {Umemoto}, T., \& {Kawabe}, R. 2003,
  \apj, 584, 357

\bibitem[{{Khanzadyan} {et~al.}(2004){Khanzadyan}, {Gredel}, {Smith}, \&
  {Stanke}}]{kha04}
{Khanzadyan}, T., {Gredel}, R., {Smith}, M.~D., \& {Stanke}, T. 2004, \aap,
  426, 171

\bibitem[{{Kumar} {et~al.}(2007){Kumar}, {Davis}, {Grave}, {Ferreira}, \&
  {Froebrich}}]{kumar07}
{Kumar}, M.~S.~N., {Davis}, C.~J., {Grave}, J.~M.~C., {Ferreira}, B., \&
  {Froebrich}, D. 2007, \mnras, 374, 54

\bibitem[{{Lee} {et~al.}(2002){Lee}, {Mundy}, {Stone}, \& {Ostriker}}]{lee02}
{Lee}, C.-F., {Mundy}, L.~G., {Stone}, J.~M., \& {Ostriker}, E.~C. 2002, \apj,
  576, 294

\bibitem[{{Leous} {et~al.}(1991){Leous}, {Feigelson}, {Andre}, \&
  {Montmerle}}]{lfam}
{Leous}, J.~A., {Feigelson}, E.~D., {Andre}, P., \& {Montmerle}, T. 1991, \apj,
  379, 683

\bibitem[{{Loinard} {et~al.}(2008){Loinard}, {Torres}, {Mioduszewski}, \&
  {Rodr{\'{\i}}guez}}]{loinard08}
{Loinard}, L., {Torres}, R.~M., {Mioduszewski}, A.~J., \& {Rodr{\'{\i}}guez},
  L.~F. 2008, \apjl, 675, L29

\bibitem[{{Lombardi} {et~al.}(2008){Lombardi}, {Lada}, \& {Alves}}]{lombardi08}
{Lombardi}, M., {Lada}, C.~J., \& {Alves}, J. 2008, \aap, 480, 785

\bibitem[{{Loren}(1989{\natexlab{a}})}]{loren89a}
{Loren}, R.~B. 1989{\natexlab{a}}, \apj, 338, 902

\bibitem[{{Loren}(1989{\natexlab{b}})}]{loren89b}
{Loren}, R.~B. 1989{\natexlab{b}}, \apj, 338, 925

\bibitem[{{Lucas} {et~al.}(2008){Lucas}, {Hoare}, {Longmore}, {Schr{\"o}der},
  {Davis}, {Adamson}, {Bandyopadhyay}, {de Grijs}, {Smith}, {Gosling},
  {Mitchison}, {G{\'a}sp{\'a}r}, {Coe}, {Tamura}, {Parker}, {Irwin}, {Hambly},
  {Bryant}, {Collins}, {Cross}, {Evans}, {Gonzalez-Solares}, {Hodgkin},
  {Lewis}, {Read}, {Riello}, {Sutorius}, {Lawrence}, {Drew}, {Dye}, \&
  {Thompson}}]{lucas08}
{Lucas}, P.~W., {Hoare}, M.~G., {Longmore}, A., {et~al.} 2008, \mnras, 391, 136

\bibitem[{{Lynds}(1962)}]{lynds62}
{Lynds}, B.~T. 1962, \apjs, 7, 1

\bibitem[{{Makarov}(2007)}]{makarov07}
{Makarov}, V.~V. 2007, \apj, 670, 1225

\bibitem[{Mamajek(2008)}]{mamajek08}
Mamajek, E.~E. 2008, Astronomische Nachrichten, 329, 10

\bibitem[{{McClure} {et~al.}(2010){McClure}, {Furlan}, {Manoj}, {Luhman},
  {Watson}, {Forrest}, {Espaillat}, {Calvet}, {D'Alessio}, {Sargent}, {Tobin},
  \& {Chiang}}]{irs10}
{McClure}, M.~K., {Furlan}, E., {Manoj}, P., {et~al.} 2010, \apjs, 188, 75

\bibitem[{{McGroarty} {et~al.}(2007){McGroarty}, {Ray}, \& {Froebrich}}]{mcg07}
{McGroarty}, F., {Ray}, T.~P., \& {Froebrich}, D. 2007, \aap, 467, 1197

\bibitem[{{Menten} {et~al.}(2007){Menten}, {Reid}, {Forbrich}, \&
  {Brunthaler}}]{menten07}
{Menten}, K.~M., {Reid}, M.~J., {Forbrich}, J., \& {Brunthaler}, A. 2007, \aap,
  474, 515

\bibitem[{{Moorwood} {et~al.}(1998){Moorwood}, {Cuby}, \& {Lidman}}]{sofi}
{Moorwood}, A., {Cuby}, J.-G., \& {Lidman}, C. 1998, The Messenger, 91, 9

\bibitem[{{Motte} {et~al.}(1998){Motte}, {Andre}, \& {Neri}}]{man98}
{Motte}, F., {Andre}, P., \& {Neri}, R. 1998, \aap, 336, 150

\bibitem[{{Mouschovias}(1976)}]{mous76}
{Mouschovias}, T.~C. 1976, \apj, 207, 141

\bibitem[{{Nakamura} {et~al.}(2011){Nakamura}, {Kamada}, {Kamazaki}, {Kawabe},
  {Kitamura}, {Shimajiri}, {Tsukagoshi}, {Tachihara}, {Akashi}, {Azegami},
  {Ikeda}, {Kurono}, {Li}, {Miura}, {Nishi}, \& {Umemoto}}]{naka11}
{Nakamura}, F., {Kamada}, Y., {Kamazaki}, T., {et~al.} 2011, \apj, 726, 46

\bibitem[{{Ozawa} {et~al.}(2005){Ozawa}, {Grosso}, \& {Montmerle}}]{ozawa05}
{Ozawa}, H., {Grosso}, N., \& {Montmerle}, T. 2005, \aap, 429, 963

\bibitem[{{Padgett} {et~al.}(2008){Padgett}, {Rebull}, {Stapelfeldt},
  {Chapman}, {Lai}, {Mundy}, {Evans}, {Brooke}, {Cieza}, {Spiesman},
  {Noriega-Crespo}, {McCabe}, {Allen}, {Blake}, {Harvey}, {Huard},
  {J{\o}rgensen}, {Koerner}, {Myers}, {Sargent}, {Teuben}, {van Dishoeck},
  {Wahhaj}, \& {Young}}]{pad08}
{Padgett}, D.~L., {Rebull}, L.~M., {Stapelfeldt}, K.~R., {et~al.} 2008, \apj,
  672, 1013

\bibitem[{{Phelps} \& {Barsony}(2004)}]{phe04}
{Phelps}, R.~L. \& {Barsony}, M. 2004, \aj, 127, 420

\bibitem[{{Pillitteri} {et~al.}(2010){Pillitteri}, {Sciortino}, {Flaccomio},
  {Stelzer}, {Micela}, {Damiani}, {Testi}, {Montmerle}, {Grosso}, {Favata}, \&
  {Giardino}}]{droxo10}
{Pillitteri}, I., {Sciortino}, S., {Flaccomio}, E., {et~al.} 2010, \aap, 519,
  A34+

\bibitem[{{Reipurth} \& {Bally}(2001)}]{rb01}
{Reipurth}, B. \& {Bally}, J. 2001, \araa, 39, 403

\bibitem[{{Ridge} {et~al.}(2006){Ridge}, {Di Francesco}, {Kirk}, {Li},
  {Goodman}, {Alves}, {Arce}, {Borkin}, {Caselli}, {Foster}, {Heyer},
  {Johnstone}, {Kosslyn}, {Lombardi}, {Pineda}, {Schnee}, \& {Tafalla}}]{com06}
{Ridge}, N.~A., {Di Francesco}, J., {Kirk}, H., {et~al.} 2006, \aj, 131, 2921

\bibitem[{{Sekimoto} {et~al.}(1997){Sekimoto}, {Tatematsu}, {Umemoto},
  {Koyama}, {Tsuboi}, {Hirano}, \& {Yamamoto}}]{sek97}
{Sekimoto}, Y., {Tatematsu}, K., {Umemoto}, T., {et~al.} 1997, \apjl, 489, L63+

\bibitem[{{Shang} {et~al.}(2007){Shang}, {Li}, \& {Hirano}}]{shang07}
{Shang}, H., {Li}, Z.-Y., \& {Hirano}, N. 2007, Protostars and Planets V, 261

\bibitem[{{Shu} {et~al.}(1987){Shu}, {Adams}, \& {Lizano}}]{shu87}
{Shu}, F.~H., {Adams}, F.~C., \& {Lizano}, S. 1987, \araa, 25, 23

\bibitem[{{Stanke}(2000)}]{phdstanke}
{Stanke}, T. 2000, PhD thesis, PhD Thesis, Mathematisch-Naturwissenschaftlichen
  Fakult{\"a}t der Universit{\"a}t Potsdam

\bibitem[{{Stanke} {et~al.}(2002){Stanke}, {McCaughrean}, \&
  {Zinnecker}}]{stanke02}
{Stanke}, T., {McCaughrean}, M.~J., \& {Zinnecker}, H. 2002, \aap, 392, 239

\bibitem[{{Stanke} {et~al.}(2006){Stanke}, {Smith}, {Gredel}, \&
  {Khanzadyan}}]{stanke06}
{Stanke}, T., {Smith}, M.~D., {Gredel}, R., \& {Khanzadyan}, T. 2006, \aap,
  447, 609

\bibitem[{{Struve} \& {Rudkj{\o}bing}(1949)}]{emsr}
{Struve}, O. \& {Rudkj{\o}bing}, M. 1949, \apj, 109, 92

\bibitem[{{Terebey} {et~al.}(1989){Terebey}, {Vogel}, \& {Myers}}]{tere89}
{Terebey}, S., {Vogel}, S.~N., \& {Myers}, P.~C. 1989, \apj, 340, 472

\bibitem[{Walawender {et~al.}(2005)Walawender, Bally, \& Reipurth}]{hhperseus}
Walawender, J., Bally, J., \& Reipurth, B. 2005, The Astronomical Journal, 129,
  2308

\bibitem[{{Wang} \& {Henning}(2006)}]{wang06}
{Wang}, H. \& {Henning}, T. 2006, \apj, 643, 985

\bibitem[{{Wang} \& {Henning}(2009)}]{wang09}
{Wang}, H. \& {Henning}, T. 2009, \aj, 138, 1072

\bibitem[{{Wang} {et~al.}(2004){Wang}, {Mundt}, {Henning}, \& {Apai}}]{wang04}
{Wang}, H., {Mundt}, R., {Henning}, T., \& {Apai}, D. 2004, \apj, 617, 1191

\bibitem[{{Wang} {et~al.}(2005){Wang}, {Stecklum}, \& {Henning}}]{wang05}
{Wang}, H., {Stecklum}, B., \& {Henning}, T. 2005, \aap, 437, 169

\bibitem[{{Wilking} {et~al.}(2001){Wilking}, {Bontemps}, {Schuler}, {Greene},
  \& {Andr{\'e}}}]{wilking01}
{Wilking}, B.~A., {Bontemps}, S., {Schuler}, R.~E., {Greene}, T.~P., \&
  {Andr{\'e}}, P. 2001, \apj, 551, 357

\bibitem[{{Wilking} {et~al.}(2008){Wilking}, {Gagn{\'e}}, \& {Allen}}]{book08}
{Wilking}, B.~A., {Gagn{\'e}}, M., \& {Allen}, L.~E. 2008, {Star Formation in
  the {$\rho$} Ophiuchi Molecular Cloud}, ed. {Reipurth, B.}, 351--+

\bibitem[{{Wilking} \& {Lada}(1983)}]{wl}
{Wilking}, B.~A. \& {Lada}, C.~J. 1983, \apj, 274, 698

\bibitem[{{Wilking} {et~al.}(1989){Wilking}, {Lada}, \& {Young}}]{wly}
{Wilking}, B.~A., {Lada}, C.~J., \& {Young}, E.~T. 1989, \apj, 340, 823

\bibitem[{{Wilking} {et~al.}(2005){Wilking}, {Meyer}, {Robinson}, \&
  {Greene}}]{wilking05}
{Wilking}, B.~A., {Meyer}, M.~R., {Robinson}, J.~G., \& {Greene}, T.~P. 2005,
  \aj, 130, 1733

\bibitem[{{Wilking} {et~al.}(1987{\natexlab{a}}){Wilking}, {Schwartz}, \&
  {Blackwell}}]{wilking87}
{Wilking}, B.~A., {Schwartz}, R.~D., \& {Blackwell}, J.~H. 1987{\natexlab{a}},
  \aj, 94, 106

\bibitem[{{Wilking} {et~al.}(1987{\natexlab{b}}){Wilking}, {Schwartz}, \&
  {Blackwell}}]{wsb}
{Wilking}, B.~A., {Schwartz}, R.~D., \& {Blackwell}, J.~H. 1987{\natexlab{b}},
  \aj, 94, 106

\bibitem[{{Wilking} {et~al.}(1997){Wilking}, {Schwartz}, {Fanetti}, \&
  {Friel}}]{wilking97}
{Wilking}, B.~A., {Schwartz}, R.~D., {Fanetti}, T.~M., \& {Friel}, E.~D. 1997,
  \pasp, 109, 549

\bibitem[{{Wu} {et~al.}(2002){Wu}, {Wang}, {Yang}, {Deng}, \& {Chen}}]{wu02}
{Wu}, J., {Wang}, M., {Yang}, J., {Deng}, L., \& {Chen}, J. 2002, \aj, 123,
  1986

\bibitem[{{Wu} {et~al.}(2004){Wu}, {Wei}, {Zhao}, {Shi}, {Yu}, {Qin}, \&
  {Huang}}]{wu04}
{Wu}, Y., {Wei}, Y., {Zhao}, M., {et~al.} 2004, \aap, 426, 503

\bibitem[{{Ybarra} {et~al.}(2006){Ybarra}, {Barsony}, {Haisch}, {Jarrett},
  {Sahai}, \& {Weinberger}}]{yba06}
{Ybarra}, J.~E., {Barsony}, M., {Haisch}, Jr., K.~E., {et~al.} 2006, \apjl,
  647, L159

\bibitem[{{Young} {et~al.}(1986{\natexlab{a}}){Young}, {Lada}, \&
  {Wilking}}]{young86}
{Young}, E.~T., {Lada}, C.~J., \& {Wilking}, B.~A. 1986{\natexlab{a}}, \apjl,
  304, L45

\bibitem[{{Young} {et~al.}(1986{\natexlab{b}}){Young}, {Lada}, \&
  {Wilking}}]{ylw}
{Young}, E.~T., {Lada}, C.~J., \& {Wilking}, B.~A. 1986{\natexlab{b}}, \apjl,
  304, L45

\bibitem[{{Young} {et~al.}(2006){Young}, {Enoch}, {Evans}, {Glenn}, {Sargent},
  {Huard}, {Aguirre}, {Golwala}, {Haig}, {Harvey}, {Laurent}, {Mauskopf}, \&
  {Sayers}}]{young06}
{Young}, K.~E., {Enoch}, M.~L., {Evans}, II, N.~J., {et~al.} 2006, \apj, 644,
  326

\bibitem[{{Zhang} \& {Wang}(2009)}]{zw09}
{Zhang}, M. \& {Wang}, H. 2009, \aj, 138, 1830

\end{thebibliography}

\clearpage

\Online
\appendix
\section{Description of the H$_{2}$ outflows in Ophiuchus}
We detected \mybf{six} new MHOs and \mybf{32} known MHOs, with \mybf{107} MHO features in total. We obtained the proper motion measurements for 86 MHO features. Table~\ref{mhopm} lists the information of each MHO feature.

In this section, we give a brief description of each MHO. Here we offer typically four images for each MHO, one H$_{2}$ image with the MHO and YSO identifications, one H$_{2}$ image with the proper motion measurements marked, one Ks image for comparison and one continuum-subtracted H$_{2}$ image to emphasize the MHO identifications.

\begin{figure*}[htb]
\centering
\includegraphics[scale=1.0]{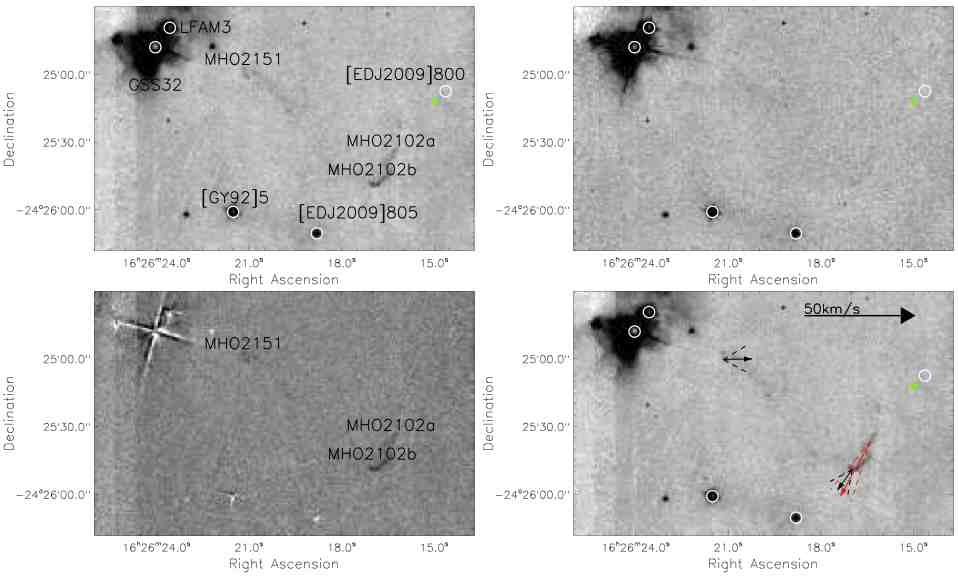}
\caption{Region of MHO 2102 and MHO 2151. \textit{Top-left}: The H$_{2}$ image with MHOs and YSOs labeled; \textit{Top-right}: The Ks image for the same region; \textit{Bottom-left}: The continuum-subtracted H$_{2}$ image with MHOs labeled; \textit{Bottom-right}: The H$_{2}$ image with the PMs marked. The YSOs from \citet{evans09} are marked with white or black circles and the X-ray sources from \citet{droxo10} are labeled with red crosses. The positions of the sub-millimeter sources from \citet{johnstone2000} are marked with green filled diamonds and the positions of the millimeter sources from \citet{young06} with blue filled squares. The PM vectors of MHOs are shown with arrows and the dashed lines show the errors of PA. The black arrows represent reliable PMs and the red arrows represent uncertain PMs. Source designations used are as follows: [EDJ2009] is from \citet{evans09}, [GY92] is from \citet{gy92}, GSS is from \citet{gss}, LFAM is from \citet{lfam}, YLW is from \citet{ylw}, BBRCG is from \citet{bbrcg}, WLY is from \citet{wly}, WSB is from \citet{wsb}, WL is from \citet{wl}, and EM*SR is from \citet{emsr}.}
\label{figA1003n}
\end{figure*}

\begin{figure*}[htb]
\centering
\includegraphics[scale=1.0]{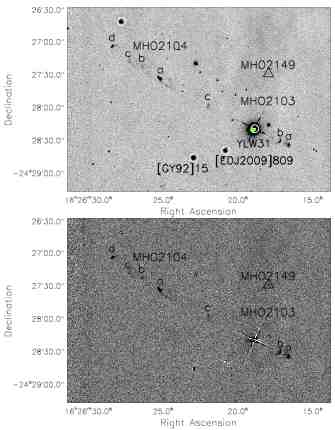}
\caption{Region of MHO 2103-2104 and MHO 2149. \textit{Top}: The H$_{2}$ 2.12 $\mu$m image with MHOs and YSOs labeled; \textit{Bottom}: The continuum-subtracted H$_{2}$ image with MHOs labeled. Others are the same as Fig.~\ref{figA1003n}. Note that there are no PM measurements for MHO 2103-2104 and MHO 2149 due to the lack of epoch1 imaging.}
\label{figA1001}
\end{figure*}

\begin{figure*}[htb]
\centering
\includegraphics[scale=1.0]{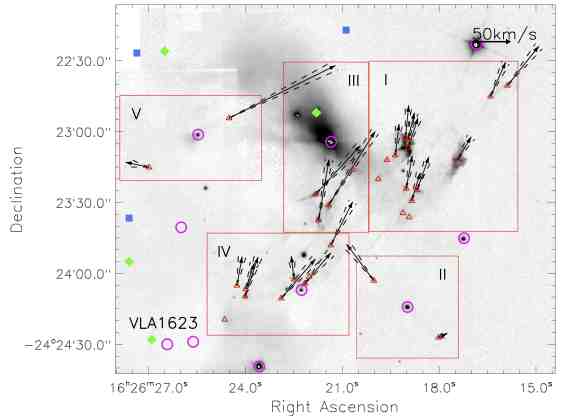}
\caption{Large view of the region of VLA 1623. The background is the H$_{2}$ 2.12 $\mu$m image. Red boxes I-V are zoomed in following figures: Fig.~\ref{figApI}, \ref{figApIII}, \ref{figApIV}, \ref{figApV}, and \ref{figApII}. The MHOs in this region are marked with red triangles and the YSOs are labeled with pink circles. Others are the same as Fig.~\ref{figA1003n}.}
\label{figA1003main}
\end{figure*}

\begin{figure*}[htb]
\centering
\includegraphics[scale=1.0]{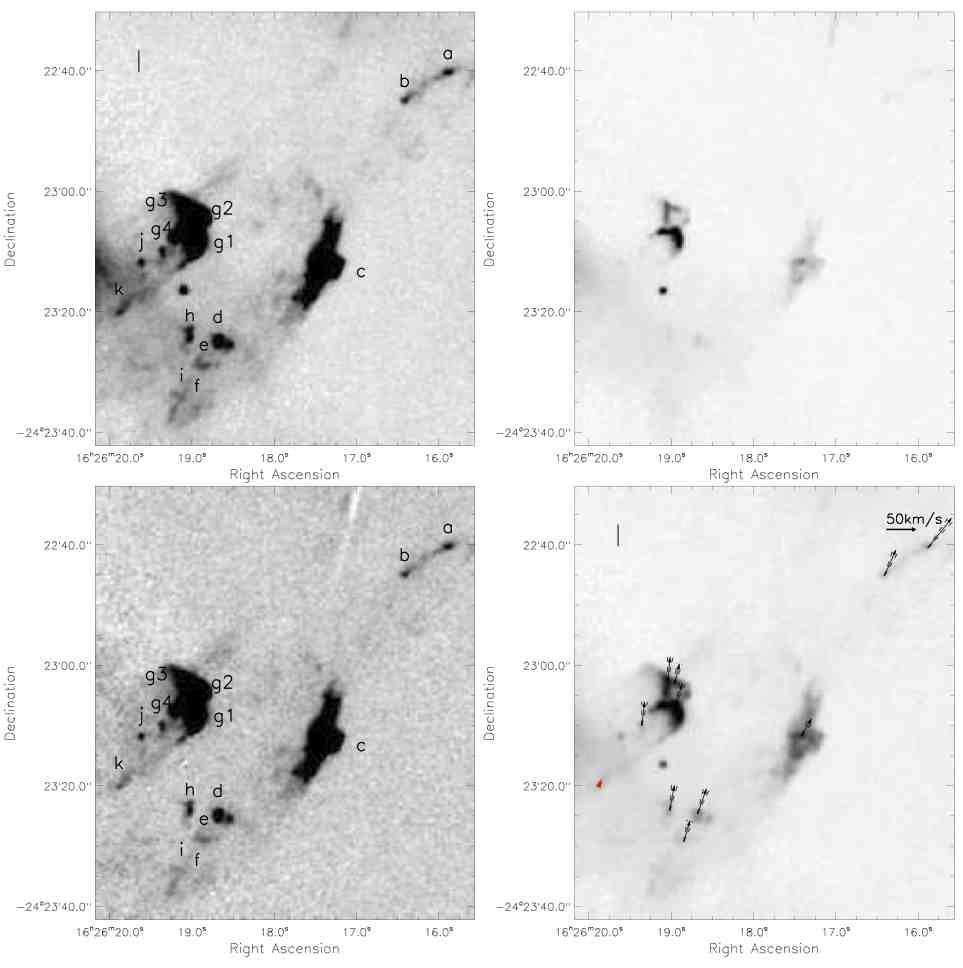}
\caption{Zoom-in view of box I in Fig.~\ref{figA1003main}. Others are the same as Fig.~\ref{figA1003n}.}
\label{figApI}
\end{figure*}

\begin{figure*}[htb]
\centering
\includegraphics[scale=1.0]{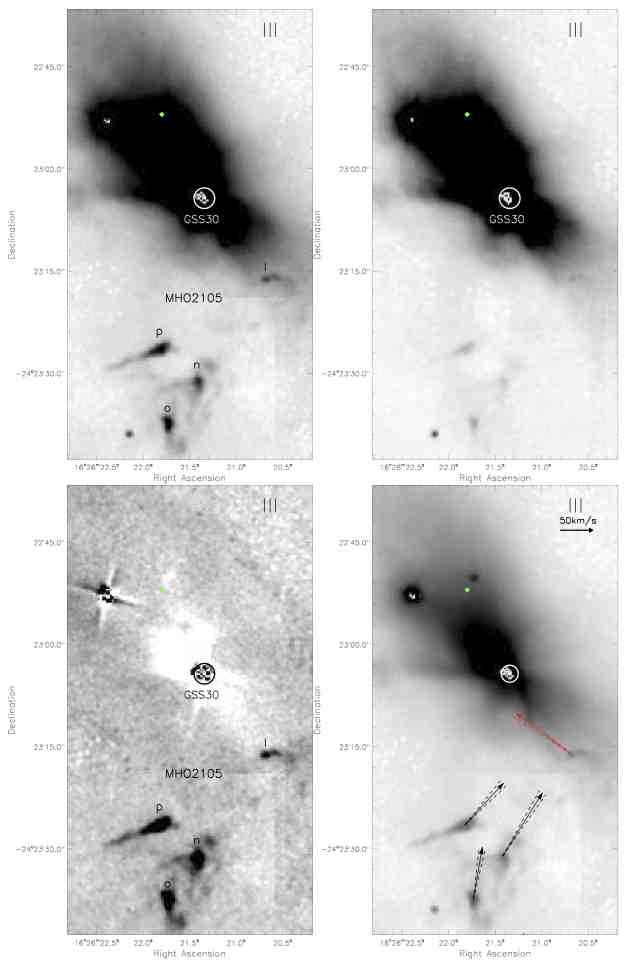}
\caption{Zoom-in view of box III in Fig.~\ref{figA1003main}. Others are the same as Fig.~\ref{figA1003n}.}
\label{figApIII}
\end{figure*}

\begin{figure*}[htb]
\centering
\includegraphics[scale=1.0]{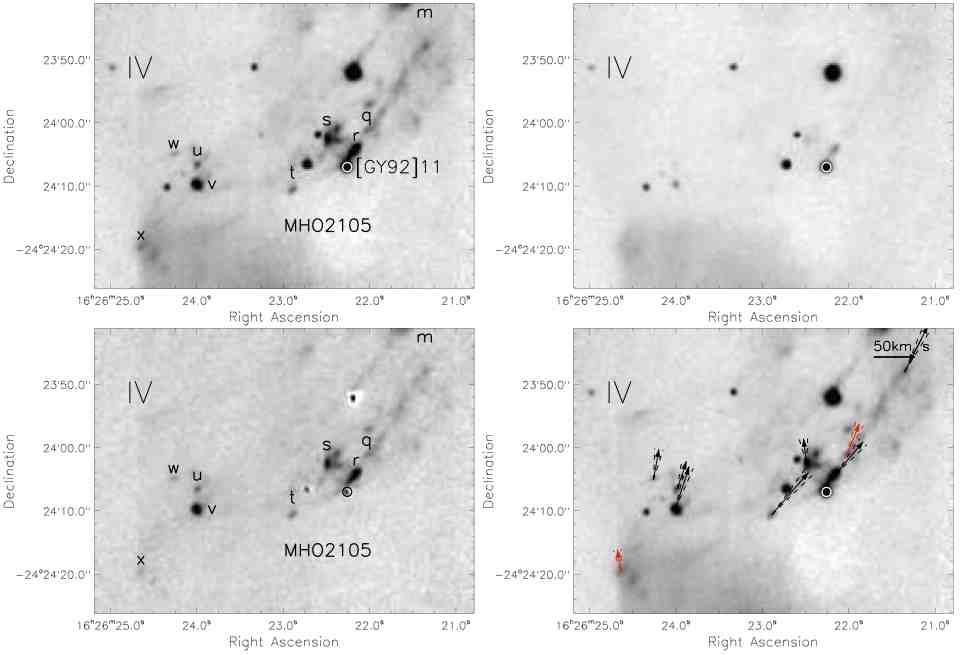}
\caption{Zoom-in view of box IV in Fig.~\ref{figA1003main}. Others are the same as Fig.~\ref{figA1003n}.}
\label{figApIV}
\end{figure*}

\begin{figure*}[htb]
\centering
\includegraphics[scale=1.0]{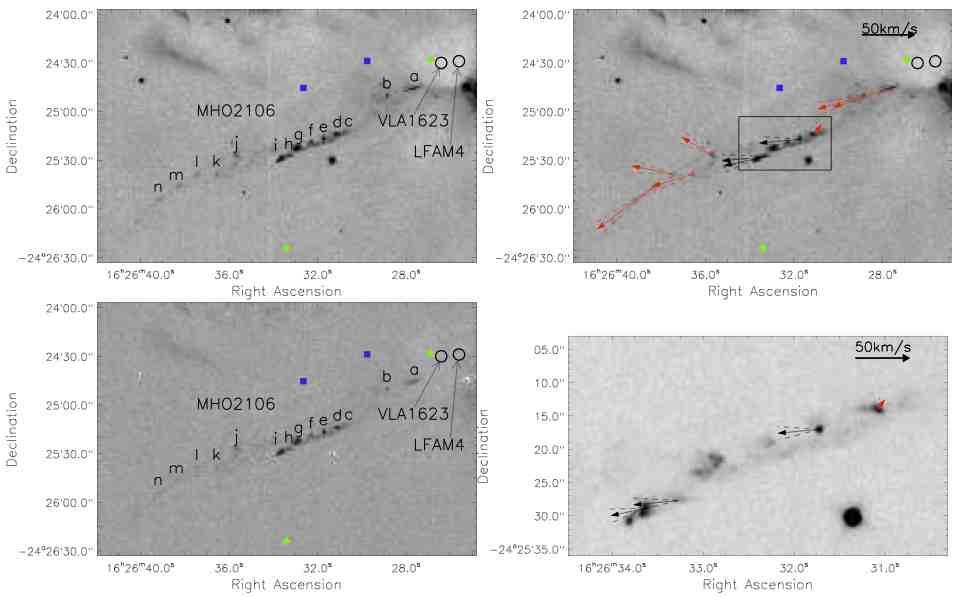}
\caption{Region of MHO 2106. \textit{Top-left} shows the H$_{2}$ image with MHOs and YSOs being labeled and \textit{Bottom-left} shows the continuum-subtracted H$_{2}$ image with the identifications of MHOs and YSOs. \textit{Top-right} shows proper motions of the features of MHO 2106 on the H$_{2}$ 2.12 $\mu$m image. \textit{Bottom-right} is a zoom-in view of the box in the top-right panel to emphasize the bright features in MHO 2106. Others are the same as Fig.~\ref{figA1003n}.}
\label{figA1004}
\end{figure*}

\begin{itemize}
\item \textbf{\textit{MHO 2102}} : The region of MHO 2102 is shown in figure~\ref{figA1003n}. This MHO consists of two knots, MHO 2102a and MHO 2102b. \citet{gome03} firstly identified this object in their H$_{2}$ narrowband images as [GSWC2003] 19 and \citet{kha04} associated it with HH 79 and HH 711 to the north according to the bow-like morphology of MHO 2102b. We obtained the proper motion (PM) measurements of this object. However, due to the low signal-to-noise ratio (SNR) of MHO 2102a in the first epoch image, the PM of MHO 2102a is uncertain. The PM vectors of MHO 2102 point back to the YSO [EDJ2009] 800 identified by \citet{evans09}. \citet{johnstone2000} also detected a sub-millimeter source about 7\arcsec~ to the southeast of [EDJ2009] 800. Based on the measured proper motions, we suggest [EDJ2009] 800 as the likely driving source of MHO 2102.

\mybf{\item \textbf{\textit{MHO 2103}} : The region of MHO 2103 is shown in figure~\ref{figA1001}. This object was firstly detected by \citet{kha04}, who identified five H$_2$ emission line knots on H$_2$ narrowband images. In our \textit{SofI} (2007) images, we detected three H$_2$ emission knots, including one new feature, MHO 2103c. MHO 2103a and b correspond to the previous detections of [KGS2004] f10-01e and f, respectively. Note that [KGS2004] f10-01a,b are located outside the coverage of our H$_{2}$ images and [KGS2004] f10-01c,g are too faint to show up in our H$_{2}$ images. We rejected [KGS2004] f10-01d as H$_{2}$ emission feature because it is invisible on our continuum-subtracted H$_{2}$ image, and we suggest it might actually be two barely resolved stars based on their morphologies in the Ks image. We did not obtain PM measurements for MHO 2103 because of the lack of previous epoch \textit{SofI} data. \citet{kha04} proposed YLW 31 as the possible driving source due to the accurate alignment of the MHO 2103 knots together with MHO 2104 knots (see the description about MHO 2104) with respect to YLW 31. We note that MHO 2103b shows a nice bow shock morphology. Based on the alignment and morphology of the MHO 2103 features, we accepted the suggestion from \citet{kha04} that YLW 31 is the driving source of MHO 2103.}

\mybf{\item \textbf{\textit{MHO 2104}} :  Figure~\ref{figA1001} shows MHO 2104 and YSOs in the region. This object was also firstly detected by \citet{kha04}, who identified two H$_{2}$ emission line knots on H$_{2}$ narrowband images. In our \textit{SofI} (2007) images, we detected four H$_{2}$ emission knots, including two new features, MHO 2104b and c. MHO 2104a and d correspond to the previous detections of [KGS2004] f10-01h and i, respectively.  We did not obtain PM measurements for MHO 2104 because of the lack of previous epoch \textit{SofI} data. \citet{kha04} proposed YLW 31 as the possible driving source due to the accurate alignment of the MHO 2104 knots with respect to YLW 31. Based on the alignment and morphology of the MHO 2104 features, we agree with the identification by \citet{kha04} that MHO 2104 is driven by YLW 31.}

\item \textbf{\textit{MHO 2105}} : Figure~\ref{figA1003main} shows the region of MHO 2105. We detected 27 H$_{2}$ emission features in MHO 2105 including six newly detected features (located in the red boxes II, III and IV in figure~\ref{figA1003main}). We also marked the position of the well-known Class 0 source VLA 1623 in Figure~\ref{figA1003main}, which is the driving source of a well-collimated high-velocity CO molecular outflow \citep{and90,dent95,naka11}. MHO 2105 is associated with the red-shifted lobe of the CO molecular outflow. \citet{davis95} and \citet{dent95} discussed in detail the association between the CO molecular outflow and the H$_{2}$ emission features in this region. \citet{gome03} suggested VLA 1623 as the driving source of most H$_{2}$ features in MHO 2105. \citet{lucas08} also detected part of the features in MHO 2105 using their UKIRT H$_{2}$ images \citep[see B and D in Fig. 19 of][]{lucas08}. 
In this paper, we obtained the PM measurements for 24 features in MHO 2105 and the results of the PM analysis support the conclusion of \citet{gome03}. Figure~\ref{figApI}, \ref{figApIII} and \ref{figApIV} show the zoomed in images for the red boxes in figure~\ref{figA1003main}. Note that for MHO 2105c,d,g,h,o,v, \citet{cog06} derived the PMs by comparing their H$_{2}$ images (March 2003) with those published by \citet{davis95} (April 1993). Our PM measurements agree with their calculations with the exception of MHO 2105c. The position angle of MHO 2105c calculated by \citet{cog06} is about 243 degrees. \citet{cog06} suggested that the driving source of MHO 2105c might be another YSO located northeast of MHO 2105c rather than VLA 1623. However, our PM measurement for MHO 2105c based on the higher resolution images gives the position angle of 329 degrees, which supports the conclusion that VLA 1623 is the driving source of MHO 2105.

\item \textbf{\textit{MHO 2106}} : Figure~\ref{figA1004} shows the region of MHO 2106. \citet{dent95} firstly detected three bright knots in MHO 2106 in their H$_{2}$ images. \citet{gome03} identified 10 features of MHO 2106 using the deeper observations. \citet{kha04} also investigated this region, increasing the number of the detected features in MHO 2106 up to 13. \citet{lucas08} detected some features based on UKIRT H$_{2}$ images (named E in Fig. 19 of their paper). Our observations of this region are the deepest to date. We confirmed 14 H$_{2}$ emission features in MHO 2106. MHO 2106 is associated with the blue-shifted lobe of the CO molecular outflow, which is driven by VLA 1623 \citep{and90,davis95,dent95,naka11}. \citet{gome03} and \citet{kha04} both suggested VLA 1623 as the driving source of MHO 2106 based on the morphology and the association with the CO outflow. We obtained the PM measurements for \mybf{10} knots in MHO 2106. However, the PM measurements for \mybf{7} faint features in MHO 2106 are uncertain due to the low SNRs for these features. The PMs of the \mybf{three} bright knots support the conclusion that MHO 2106 is driven by VLA 1623.

\item \textbf{\textit{MHO 2108}} : Figure~\ref{figA0401} shows the region of MHO 2108. \citet{kha04} detected this object as a bow shape structure in their H$_{2}$ images. They suggested a YSO northeast of MHO 2108 as the driving source. Our deeper observations reveal a more complicated structure for MHO 2108. It consists of a bow-like structure, a faint knot, and a long tail connected to the infrared nebular around WL 18. The optical counterpart of MHO 2108 is HH 673 \citep{phe04}. \citet{phe04} also detected HH 673b $\sim$ 40\arcsec southeast of WL 18. They suggested that HH 673 and HH 673b belong to a bipolar outflow driven by WL 18. We obtained a PM measurement for MHO 2108, but the PM is uncertain because of the difference of morphology between the images in the two epochs. Based on the morphology of MHO 2108, we accepted the assumption from \citet{phe04} that WL 18 is the possible driving source of MHO 2108.

\item \textbf{\textit{MHO 2109}} : Figure~\ref{figA0401} also shows MHO 2109, which exhibits a bow shape in our H$_{2}$ images. \citet{gome03} and \citet{kha04} both detected this object in their images. \citet{gome03} suggested that WL 15 (Elias 2-29, located 2\farcm4 to the east from MHO 2109 and situated outside of our image) as the driving source of MHO 2109. WL 15 is associated with a bipolar CO outflow \citep{bon96,sek97} and the red-shifted lobe roughly aligns with the direction of MHO 2109. \citet{kha04} confirmed the bow shape of MHO 2109 and they inferred that MHO 2108 and MHO 2109 arise from a flow connecting both knots with a flow direction to the northeast. \citet{naka11} investigated the CO outflows in L1688 using CO (J=3-2) and CO (J=1-0) mapping observations. They identified a CO outflow driven by the YSO LFAM 26. MHO 2109 is associated with the red-shifted lobe of this CO outflow. \mybf{We obtained the PM measurement of MHO 2109. We found that the PM vector of MHO 2109 points to the direction of the northwest, which indicates that the driving source should be located in the southeast of MHO 2109. There are some YSOs in the southeast region. Here we suggest YLW 5 as the likely driving source just because that YLW 5 is the nearest YSO to MHO 2109.}  

\item \textbf{\textit{MHO 2110}} : MHO 2110 consists of three knots, MHO 2110a-c. Figure~\ref{figA0301} shows the region of MHO 2110. \citet{kha04} detected this object in their H$_{2}$ images. They suggested EM*SR 24 as the driving source of MHO 2110 because MHO 2110, together with MHO 2111 (see the following description) and several HH objects (HH 224, HH 709 and HH 418, which are situated outside of our image), is aligned along a NW-SE direction passing through EM*SR 24 \citep[see.][Fig. 3]{kha04}. For MHO 2110, we only obtained the uncertain PMs due to the low SNRs of the emission features. Therefore, we cannot associate MHO 2110 with any YSO based on our PM measurements.

\item \textbf{\textit{MHO 2111}} : Figure~\ref{figA0302} shows the region of MHO 2111. MHO 2111 consists of two knots: MHO 2111a is a central knot with a diffuse nebula. MHO 2111b is an elongated knot. MHO 2111 has been detected by optical, near- and mid-infrared observations. Its optical counterpart is HH 224S. \citet{zw09} identified its mid-infrared counterpart using Spitzer/IRAC images. \citet{phe04} identified HH 224S at the location of MHO 2111. They also detected several HH objects \citep[HH 224N, HH 224NW1 and HH 224NW2, see][Fig. 12]{phe04} to the northwest of HH 224S. According to the alignment of the HH 224 complex, \citet{phe04} suggested [GY92] 193 (located $\sim$7\arcmin to the northwest and outside the boundary of figure~\ref{figA0302}) as the driving source of HH 224S. \citet{kha04} detected MHO 2111 at H$_{2}$ 2.12$\mu$m as the counterpart of HH 224S, but they did not detect any near-infrared counterparts of HH 224N, HH 224NW1, and HH 224NW2. Therefore they suggested that MHO 2111 might be driven by EM*SR 24, as MHO 2111, MHO 2110 and EM*SR 24 are aligned roughly along a line. However, \citet{zw09} suggested WSB49, the YSO closest to MHO 2111, as the driving source based on MHO 2111's wide-cavity lobe morphology, which faces away from WSB 49. We obtained the PM measurements for MHO 2111, found that the PM vectors point back to WSB 49. Therefore, we confirmed the assumption from \citet{zw09} that WSB 49 is the likely driving source of MHO 2111.

\item \textbf{\textit{MHO 2112}} : MHO 2112 is a faint knot (see Figure~\ref{figA0402}). \citet{kha04} detected MHO 2112 in their H$_{2}$ images. They suggested that MHO 2112 is driven by YLW 16 based on the morphology and distribution of MHO features \citep[see][Fig. 5]{kha04}. MHO 2112 is too faint to measure the proper motion. We have no enough evidence to associate MHO 2112 with the nearby YSOs.

\item \textbf{\textit{MHO 2113}} : MHO 2113 is a bright knot (see Figure~\ref{figA0402}). \citet{gome03} and \citet{kha04} both detected MHO 2113. \citet{gome03} suggested that the nearest YSO [GY92] 235 as the driving source. Based on the morphology and the location of H$_{2}$ features \citep[see][Fig. 5]{kha04}, \citet{kha04} inferred that MHO 2113 may be part of the large H$_{2}$ outflow driven by YLW 15. We obtained the PM measurement for MHO 2113. \mybf{The PM vector of MHO 2113 indicates that the driving source should be located in the north of MHO 2113. There are several YSOs in the north of MHO 2113. There is no strong evidence to associate MHO 2113 with the certain one of them. Therefore, here we accept the suggestion from \citet{gome03} that [GY92] 235 is the likely driving source of MHO 2113.}

\item \textbf{\textit{MHO 2114}} : MHO 2114 is a knot (see Figure~\ref{figA0402}). \citet{kha04} identified this object in their H$_{2}$ images. They proposed an association between MHO 2114 and the nearby YSO, [GY92] 235. \mybf{The PM of MHO 2114 obtained by us indicates that the driving source should be located in the north. There are several YSOs in the north of MHO 2114, including [GY92] 235 and [GY92] 232. We have associated MHO 2113 with [GY92] 235 and it seems impossible that MHO 2113 and MHO 2114 constitute a bipolar outflow which is driven by [GY92] 235 because MHO 2113 and MHO 2114 are located in the same side of [GY92] 235. Therefore, here we suggest another possibility that MHO 2114 is driven by [GY92] 232 because the PM vector of MHO 2114 points back to [GY92] 232.} 

\item \textbf{\textit{MHO 2115}} : Figure~\ref{figA0901} shows the region of MHO 2115. \citet{gome03} detected six H$_{2}$ features in MHO 2115, corresponding to MHO 2115a-e,g. \citet{kha04} identified an additional feature, MHO 2115f, in this MHO. \citet{lucas08} also detected some features using the UKIRT H$_{2}$ images (named G in Fig. 19 of their paper). BBRCG 24 is a Class II source \citep{evans09} and is the possible driving source of a CO molecular outflow \citep{sek97}. Based on the morphology and distribution of MHO 2115 features, \citet{gome03} and \citet{kha04} both suggested that these H$_{2}$ emission knots arise from multiple, overlapping flows. The PMs of the MHO 2115 features support this conclusion. \mybf{We infer that MHO 2115b,d,e,f,g are driven by BBRCG 24 based on the position angles of the PM vectors and the fact that MHO 2115b,d,e,g are associated with the blue lobe of the CO outflow, while MHO 2115f is associated with the red lobe of the CO outflow \citep[see][Fig. 2]{sek97}. The PM vectors of MHO 2115a,c indicate that MHO 2115a,c may be driven by another YSO and their driving source should be located in the southwest of MHO 2115a,c. We associate MHO 2115a,c with [GY92] 93, which is located $\sim$9.5\arcmin to the southwest (outside the boundary of figure~\ref{figA0901}). HH 314 is located $\sim$47\arcsec to the southwest of [GY92] 93. MHO 2115a,c and HH 314 may constitute a bipolar outflow driven by [GY92] 93.} 

\item \textbf{\textit{MHO 2116}} : MHO 2116 exhibits a bow shape in Figure~\ref{figA0902k}. Based on the morphology of MHO 2116, \citet{gome03} and \citet{kha04} suggested YLW 14 as the possible driving source of MHO 2116. The PM vector of MHO 2116 points back to YLW 14, supporting this suggestion. YLW 14 also drives a CO molecular outflow \citep[see][Fig. 2]{sek97} and MHO 2116 is roughly associated with the blue-shifted lobe of this CO outflow.

\item \textbf{\textit{MHO 2117}} : MHO 2117 shows a bow-like structure in Figure~\ref{figA0902k}. \citet{gome03} suggested that MHO 2117 and MHO 2116 share the same powering star and thus belong to the same flow. \citet{kha04} also detected MHO 2117. They suggested that the driving source should lie to the southeast of MHO 2117 based on the bow shock morphology of MHO 2117. \citet{gome03} and \citet{kha04} also proposed WLY 2-48 (outside the boundary of figure~\ref{figA0902k}) as a possible driving source as this YSO is associated with a bipolar CO outflow \citep{bon96}. However, \citet{naka11} investigated the CO outflows in this region. They associated this CO outflow with Elias 2-32, which is located to the northeast of MHO 2117, rather than with WLY 2-48. Our PM measurement for MHO 2117 is uncertain as the morphology of MHO 2117 changed somewhat from 2000 to 2007. \mybf{Moreover, the orientation of the PM vector of MHO 2117 is not in agreement with its bow-like morphology. Thus we can't associate MHO 2117 with any ambient YSOs. The PM vector of MHO 2117 indicates that the possible driven source could be located in its southwest. Therefore, MHO 2116 and MHO 2117 may have some connections as suggested by \citet{gome03}.}

\item \textbf{\textit{MHO 2118}} : MHO 2118 consists of two faint knots MHO 2118a and MHO 2118b (Figure~\ref{figA0403}). \citet{kha04} suggested YLW 16, the closest YSO, as the driving source. We obtained the proper motion of MHO 2118b, but it is uncertain because MHO 2118b has the low SNR in the first epoch image. Using the Owens Valley Millimeter Interferometer \citet{tere89} detected a CO outflow driven by YLW 16. MHO 2118 is associated with the red-shifted lobe of this CO outflow. Therefore, we also suggest YLW 16 as the driving source of MHO 2118.

\item \textbf{\textit{MHO 2119}} : MHO 2119 shows a knot with a faint tail in Figure~\ref{figA0403}. \citet{kha04} detected this object and they suggested that MHO 2119 is related to MHO 2114. The PM vector of MHO 2119 points back to YLW 16. \citet{tere89} detected an extended, blue shifted lobe north of YLW 16 in the direction of MHO 2119 \citep[see][Fig. 2b]{tere89}. It seems that MHO 2119 and MHO 2118 constitute a bipolar outflow driven by YLW 16. Therefore, we suggest YLW 16 as the driving source of MHO 2119.

\item \textbf{\textit{MHO 2120}} : MHO 2120 is faint in Figure~\ref{figA0403}. Because there are many MHOs and YSOs in this region, \citet{gome03} suggested several candidates for the driving source of MHO 2120, including YLW 15 and YLW 16. Based on the morphology and distribution of MHOs in this region \citet{kha04} inferred that MHO 2120 is part of a larger outflow driven by YLW 15. We obtained an uncertain proper motion measurement for MHO 2120 because of its low SNR in both epoch images. \mybf{Nevertheless, the orientation of the PM vector indicates that the driving source of MHO 2120 should be locate to the south of MHO 2120. YLW 15 might have physical relations with MHO 2120.}

\item \textbf{\textit{MHO 2121}} : MHO 2121 is an elongated knot in Figure~\ref{figA0403}. HH 674 is its optical counterpart \citep{gome98,phe04}. As there are many YSOs around HH 674, \citet{phe04} did not offer any definitive identification of the driving source. \citet{gome03} and \citet{kha04} both detected this object in their H$_{2}$ images. The latter suggested MHO 2121 being part of a bipolar H$_{2}$ outflow driven by YLW 16 \citep[see][Fig. 5]{kha04}. \mybf{The PM vector of MHO 2121 indicates that the driving source should lie to the southwest of MHO 2121. There are many YSOs to the southwest that are located outside our images. Thus we cannot associate MHO 2121 with any ambient YSO.}

\item \textbf{\textit{MHO 2122}} : MHO 2122 is a central knot with a faint diffuse nebula in Figure~\ref{figA0403}. Based on the morphology and distribution of the MHOs in this region, \citet{kha04} suggested MHO 2122 to be part of a bipolar outflow driven by YLW 15 \citep[see][Fig. 5]{kha04}. The PM vector of MHO 2122 indicates the driving source to its north. Therefore, we suggest YLW 47, a Class II source identified by \citet{evans09}, as the driving source of MHO 2122.

\item \textbf{\textit{MHO 2123}} : Figure~\ref{figA0403} shows the region of MHO 2123, which consists of three features. MHO 2123a is an elongated knot, MHO 2123b is a faint nebula, and MHO 2123c is a knot. \citet{gome03} detected MHO 2123a in their H$_{2}$ images and named it as [GSWC2003] 7c. \citet{kha04} detected four features in MHO 2123, including [KGS2004] f04-05h, which is too faint to confirm in our H$_{2}$ images. Based on our reprocessing of \citet{kha04} data, which resulted in a non-detection of [KGS2004] f04-05h, we suspect that it might have been misclassified in the original analysis. \citet{kha04} suggested MHO 2123 as part of a bipolar H$_{2}$ outflow driven by YLW 16 \citep[see][Fig. 5]{kha04}. We obtained PM measurements for MHO 2123a,c. MHO 2123b have no enough SNR to measure the PM. The PM of MHO 2123a is uncertain due to its low SNR in the first epoch images. \mybf{Our PM measurement for MHO 2123c indicates that its driving source should lie to the northwest. We associate MHO 2123c with YLW 47, the closest YSO in this direction.}

\item \textbf{\textit{MHO 2124}} : MHO 2124 consists of two bright nebular features with MHO 2124b being more diffuse than MHO 2124a (see Figure~\ref{figA0403}). \citet{gome03} and \citet{kha04} both detected MHO 2124. \citet{kha04} proposed YLW 16 as the driving source. \mybf{The PM vectors of MHO 2124 indicate that its driving source should be located to its southwest. MHO 2124 and MHO 2121 might be part of the same outflow. There are some YSOs to the west, lying outside our images. We cannot offer a definite identification of the driving source of MHO 2124.}

\item \textbf{\textit{MHO 2125}} : MHO 2125 is an elongated knot (Figure~\ref{figA0501}). Based on the morphology of MHO 2125, which hints at a bow shock pointing towards SE or SW, \citet{kha04} suggested that MHO 2125 might arise from a NW-SE flow driven by YLW 15 or YLW 16. We obtained the PM measurement of MHO 2125, but it is uncertain because of the low SNR in the first epoch images. This uncertain PM vector points back to the direction of WLY 2-55, an X-ray source identified by \citet{droxo10}. There is also a YSO, YLW 45, located $\sim$1.4\arcmin~to the south of MHO 2125 (outside the boundary of figure~\ref{figA0501}). \citet{bon96} detected a CO molecular outflow from YLW 45, and MHO 2125 is roughly associated with the red-shifted lobe of this CO outflow.

\item \textbf{\textit{MHO 2126}} :  MHO 2126 is a bipolar outflow consisting of two knots MHO 2126a and b.  Figure~\ref{figA0502} shows that MHO 2126 is close to [GY92] 344, a ``Flat spectrum'' source \citep{evans09} with a bipolar nebular. \citet{kha04} suggested this edge-on disk system with a bipolar outflow might be the driving source for MHO 2126. Our PM measurements of MHO 2126 support this conclusion.

\item \textbf{\textit{MHO 2127}} : MHO 2127 is a bright knot in Figure~\ref{figA0501}. \citet{kha04} found that MHO 2127 is elongated with the semi-major axis pointing towards YLW 16 (see Fig.~\ref{figA0403}). We did not obtain a PM measurement for MHO 2127 although it has a point-like structure and enough SNR in both epoch images, which means that the proper motion of MHO 2127 is very small. We note that MHO 2127 is also located to the northeast of YLW 45 ($\sim$4\arcmin~and outside the boundary of figure~\ref{figA0501}) and that the red-shifted lobe of the CO outflow driven by YLW 45 \citep{bon96} extends roughly in the direction of MHO 2127.

\item \textbf{\textit{MHO 2128}} : MHO 2128 is an extended bow-like structure in Figure~\ref{figA0801}. Figure~\ref{figAp28} shows the zoom-in image. \citet{grosso01} detected MHO 2128 in H$_{2}$ 2.12 $\mu$m. They suggested that MHO 2128 belongs to two different jets emanating from two Class I protostars YLW 15 (see Fig.~\ref{figA0403}) and YLW 52. \citet{kha04} also detected MHO 2128, together with several H$_{2}$ features to the east of YLW 52 \citep[see][Fig. 8]{kha04}. They suggested all the features to be driven by YLW 52. \citet{lucas08} also detected MHO 2128 in the UKIRT H$_{2}$ images (named I in Fig. 19 of their paper). They suggested that MHO 2128 belongs to a large outflow connecting MHO 2128, MHO 2106, MHO 2105, and some other features outside the coverage of our \textit{SofI} observations. We identified six knots in MHO 2128 and obtained their proper motions. \mybf{MHO 2128a-f show the radial PMs (Figure~\ref{figAp28}), indicating that they arise from the same driving source.} The PM vectors indicate an driving source to its east. The proper motion results agree with the suggestion from \citet{kha04} and contradict the suggestions of MHO 2128 being part of a large outflow driven by a source located to its west.

\item \textbf{\textit{MHO 2129}} : MHO 2129 is a knot in Figure~\ref{figA0801}, with a zoom-in shown in Figure~\ref{figAp29}. Note that Figure~\ref{figAp29} emphasizes the identification of MHO 2129 with the inverted-gray-scale images. MHO 2129 corresponds to the identification of [KGS2004] f08-01b by \citet{kha04}. They also identified other two features, [KGS2004] f08-01c and d, in the immediate vicinity of YLW 52. However, [KGS2004] f08-01c and d are invisible in our continuum-subtracted H$_{2}$ images. We suspect that these two features are the mis-identified noise points. We obtained the PM measurement for MHO 2129, but it is uncertain because of the low SNR in both epoch images. Given the location of MHO 2129, \citet{kha04} proposed YLW 52 as the driving source.

\item \textbf{\textit{MHO 2130}} : MHO 2130 consists of two features in Figure~\ref{figA0801}, with a zoom-in shown in Figure~\ref{figAp30}. Note that Figure~\ref{figAp30} emphasizes the identification of MHO 2130 with the inverted-gray-scale images. MHO 2130a is an elongated knot and MHO 2130b is a faint nebula. \citet{kha04} suggested that MHO 2130 arises from the walls of a cavity, possibly excited by oblique shocks in an S-shape flow from YLW 52. MHO 2130b is too faint for a PM measurement, but the SNR of MHO 2130a is sufficiently high. \mybf{We obtained the PM measurement for MHO 2130a and the PM vector indicates that the driving source should be located in the southwest of MHO 2130a. Here we accept the suggestion from \citet{kha04} that MHO 2130 is driven by YLW 52.}

\item \textbf{\textit{MHO 2131}} : MHO 2131 is an elongated knot in Figure~\ref{figA0801}. MHO 2131 corresponds to the identification of [KGS2004] f08-01g by \citet{kha04}. They also detected a feature near MHO 2131, named [KGS2004] f08-01h, but [KGS2004] f08-01h is too faint in our continuum-subtracted H$_{2}$ images to be confirmed. \citet{kha04} suggested YLW 52 as the driving source of MHO 2131 and our PM measurement for MHO 2131 support this conclusion. It seems that MHO 2128-2131 constitute an S-shape H$_{2}$ outflow from YLW 52.

\item \textbf{\textit{MHO 2132}} : MHO 2132 consists of three features in Figure~\ref{figA0504}. \citet{stanke06} detected the Class 0 source MMS 126, located right between MHO 2132b and c. MMS 126 drives a CO molecular outflow \citep{stanke06} and MHO 2132 is well associated with this CO outflow. Therefore, \citet{kha04} suggested MMS 126 as the driving source of MHO 2132. Our PM measurements for MHO 2132 support this conclusion. Note that the PM of MHO 2132c is uncertain because of its low SNR in the first epoch images. Actually, MHO 2132 has a mid-infrared counterpart in the Spitzer/IRAC images \citep{zw09} and MMS 126 is also visible in the IRAC and MIPS images. The outflow shows more extended and more obvious structures in the IRAC images, including a bipolar nebular around MMS 126 and the emission to the north of MHO 2132c and to the south of MHO 2132a \citep[see][Fig. 5 and 6]{zw09}.

\item \textbf{\textit{MHO 2137}} : Figure~\ref{figA1003main} shows the region of MHO 2137, with a zoom-in shown in Figure~\ref{figApV}. \citet{gome03} detected this object, but they did not identify a driving source for this outflow. The PM vectors determined by us indicate a bipolar outflow from the Class I source [GY92] 30 \citep{evans09}, which shows up as an infrared nebula in the Ks image.

\item \textbf{\textit{MHO 2149}} : MHO 2149 is a knot shown in Figure~\ref{figA1001}. It is newly discovered in our deep H$_{2}$ images. We note that YLW 31 is located to the south of MHO 2149, but YLW 31 drives a NE-SW H$_{2}$ outflow (see the description of MHO 2104) and it has been confirmed as a single star with a resolution of 0.25\arcsec~\citep{cos00,bar03}. Therefore, MHO 2149 is unlikely to be driven by YLW 31. \citet{phe04} detected the candidate HH object O1a within about 10\arcsec~of MHO 2149. They suggested a relation between O1 and HH 419, which is located $\sim$3\arcmin~to the southwest of MHO 2149. Therefore, MHO 2149 and HH 419 may arise from the same outflow.

\item \textbf{\textit{MHO 2150}} : Figure~\ref{figA1003main} shows the region of MHO 2150, with a zoom-in shown in Figure~\ref{figApII}. MHO 2150 consists of two features MHO 2150a, a bright knot with a diffuse tail, and MHO 2150b, a dumbbell-shape knot. \citet{cog06} firstly detected this object, naming the features as A and B, but this identification is not included in the \textit{SIMBAD} database. We obtained the PMs of MHO 2150 and the PM vectors indicate that MHO 2150a and b constitute a bipolar outflow driven by the central source [EDJ2009] 807.

\item \textbf{\textit{MHO 2151}} : Figure~\ref{figA1003n} shows the region of MHO 2151. MHO 2151 is a new feature with no previous identification. We obtained the PM of MHO 2151 and the PM vector indicates that the driving source of MHO 2151 should be located to the east. \mybf{There are two YSOs in the east of MHO 2151, GSS 32 and LFAM 3. \citet{kamazaki03} detected a NW-SE CO molecular outflow named as AS outflow \citep[see Fig.5 in][]{kamazaki03} which is centered around LFAM 3. They suggested LFAM 3 as the driven source of the AS outflow because LFAM 3 is associated with the 6 cm radio continuum emission, which is considered to be an indicator of the active ionized jet, and with the 1.3 mm dust continuum source. However, we found that MHO 2151 is not associated with blue-shifted or red-shifted lobe of AS outflow, so it seems impossible that MHO 2151 is driven by LFAM 3.} Therefore, here we associate MHO 2151 with the nearby YSO GSS 32 based on the alignment of the PM vector and GSS 32. We also can see some faint extended structures between MHO 2151 and MHO 2102 (driven by [EDJ2009] 800; see the description about MHO 2102), which may indicate some interactions between these two H$_{2}$ outflows.

\item \textbf{\textit{MHO 2152}} : Figure~\ref{figA0902n} shows the region of MHO 2152. MHO 2152 is a newly discovered feature with no previous identification. We obtained the PM of MHO 2152 and the PM vector indicates that the driving source of MHO 2152 should lie to the \mybf{southwest. There are several YSOs in the southwest of MHO 2152. We have no sufficient evidence to associate MHO 2152 with one of them.} 

\item \textbf{\textit{MHO 2153}} : Figure~\ref{figA0902n} shows the region of MHO 2153. MHO 2153 is a newly discovered feature with no previous identification. We obtained the PM of MHO 2153 and the PM vector indicates that the driving source of MHO 2153 should lie to the \mybf{southwest. There are several YSOs in the southwest of MHO 2153, including [GY92] 236 and [GY92] 239. Here we suggest [GY92] 239 as the likely driven source of MHO 2153 just because [GY92] 239 is the closest YSO to MHO 2153.}

\item \textbf{\textit{MHO 2154}} : Figure~\ref{figA0504} shows the region of MHO 2154, which is a faint nebula in our H$_{2}$ image. \citet{zw09} identified a mid-infrared counterpart in the IRAC images, named EGO 33 \citep[see][Fig. 5 and 6]{zw09}. They suggested YLW 58 as the driving source of EGO 33 based on that MMS 126 has a close association with the known outflow detected in the near-infrared and in CO (3-2) emission \citep{stanke06}. They argued that EGO 33 is unlikely to be associated with MMS 126 as EGO 33 does not coincide with the axis of the known CO outflow. MHO 2154 is invisible in the first epoch images, thus we could not obtain its proper motion. Here we accept the suggestion from \citet{zw09} that YLW 58 is the driving source of MHO 2154.

\item \textbf{\textit{MHO 2155}} : MHO 2155 shows a elongated knot in Figure~\ref{figA0902k}. MHO 2155 is a newly discovered feature with no previous identification. We did not get the PM measurement for MHO 2155 because it is invisible in the first epoch images. It seems that there are some faint structures between MHO 2155 and MHO 2117 in the H$_2$ narrowband image, which indicates that there may be some physical connections between MHO 2155 and MHO 2117.

\end{itemize}

\longtabL{1}{
\begin{landscape}
\begin{center}

\footnotesize
\begin{longtable}{lcclccccc}

\caption{\label{mhopm} Proper motions of molecular hydrogen outflows}\\
\hline\hline
MHO & RA \tablefootmark{a}& Dec.\tablefootmark{a} & Other names & $\mu_{\alpha}$cos$\delta$ & $\mu_{\delta}$ & PA\tablefootmark{b} & Velocity\tablefootmark{c} &Status\tablefootmark{d} \\
 & (J2000) & (J2000) & & (mas/yr) & (mas/yr) & (deg) & (km/s) & Flag\\
\hline
\endfirsthead
\caption{continued.}\\
\hline\hline
MHO& RA\tablefootmark{a} & Dec.\tablefootmark{a} & Other names & $\mu_{\alpha}cos\delta$ & $\mu_{\delta}$ & PA\tablefootmark{b} & Velocity \tablefootmark{c}&Status\tablefootmark{d} \\
 & (J2000) & (J2000) & & (mas/yr) & (mas/yr) & (deg) & (km/s) & Flag\\
\hline
\endhead
\hline
\endfoot
MHO2102a & 16 26 16.4 &-24 25 37& [KGS2004]f10-03a & 21.9$\pm$ 7.7& -39.9$\pm$ 10.9&151$\pm$10& 25.7$\pm$ 5.9& 2\\                                                                                                                                                                                                                                                                                                                                                                                                  
MHO2102b & 16 26 17.0 &-24 25 49& [GSWC2003]19,f10-03b & 10.6$\pm$ 7.6& -14.8$\pm$ 9.6&144$\pm$26& 10.3$\pm$ 5.1& 0\\                                                                                                                                                                                                                                                                                                                                                                                               
MHO2103a & 16 26 16.6 &-24 28 34& [KGS2004]f10-01e & \ldots& \ldots&\ldots& \ldots& \ldots\\                                                                                                                                                                                                                                                                                                                                                                                                                        
MHO2103b & 16 26 17.1 &-24 28 31& [KGS2004]f10-01f & \ldots& \ldots&\ldots& \ldots& \ldots\\
\end{longtable}
\end{center}
\tablefoottext{a}{The position of brightness peak within the polygon defined in the \textit{SofI} (2007) H$_{2}$ image according to the surface brightness distribution of MHO feature.}
\tablefoottext{b}{Position angle of PM vector measured east of north.}
\tablefoottext{c}{Tangential velocity is calculated assuming that the MHO feature is located at a distance of 119 pc.}
\tablefoottext{d}{Status flag. Value 0 represents reliable PM measurements. Value 2 represents uncertain PM measurements, whose uncertainty is mainly due to low SNR in one or two epoch images. Value 1 also represents uncertain PM measurements, whose uncertainty is mainly due to the change in morphology.} 
\tablefoottext{e}{These names are from \citet{cog06} and they are not included in the \textit{SIMBAD} database.}

\end{landscape}
}
\clearpage

\clearpage

\clearpage

\clearpage

\begin{figure*}[htb]
\centering
\includegraphics[scale=1.0]{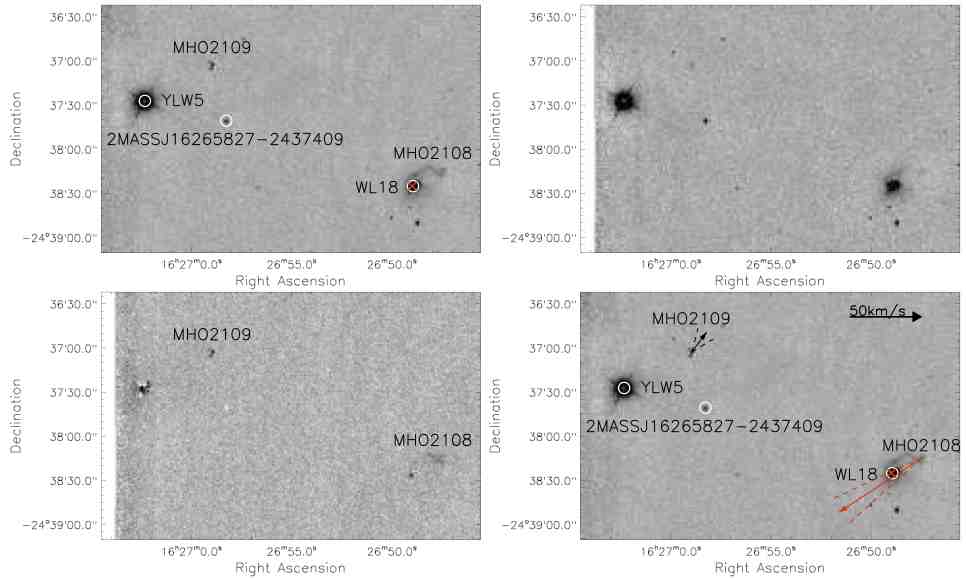}
\caption{Same as Fig.~\ref{figA1003n}, but for the region of MHO 2108 and MHO 2109.}
\label{figA0401}
\end{figure*}

\begin{figure*}[htb]
\centering
\includegraphics[scale=1.0]{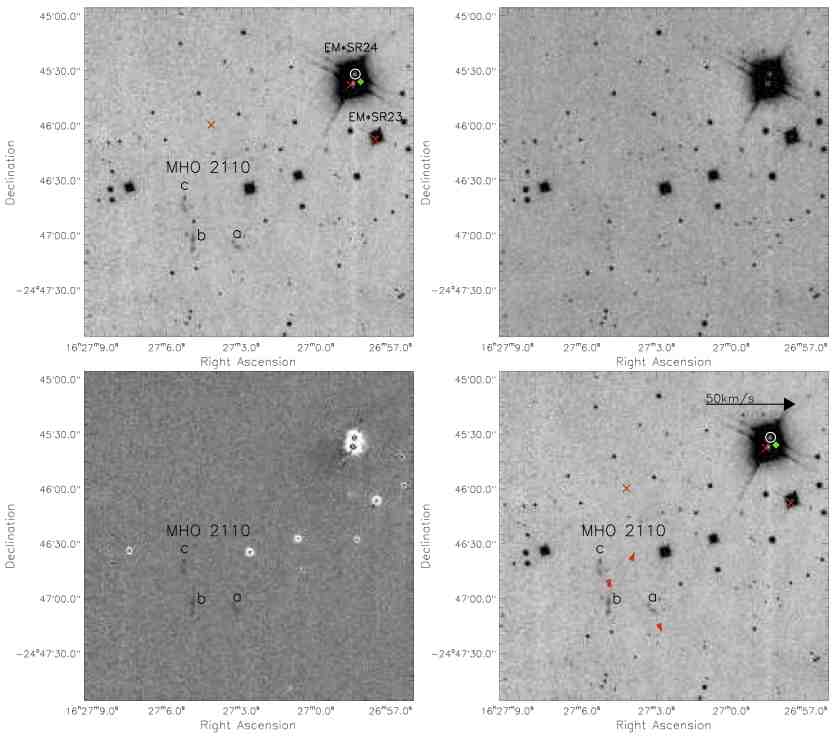}
\caption{Same as Fig.~\ref{figA1003n}, but for the region of MHO 2110.}
\label{figA0301}
\end{figure*}

\begin{figure*}[htb]
\centering
\includegraphics[scale=1.0]{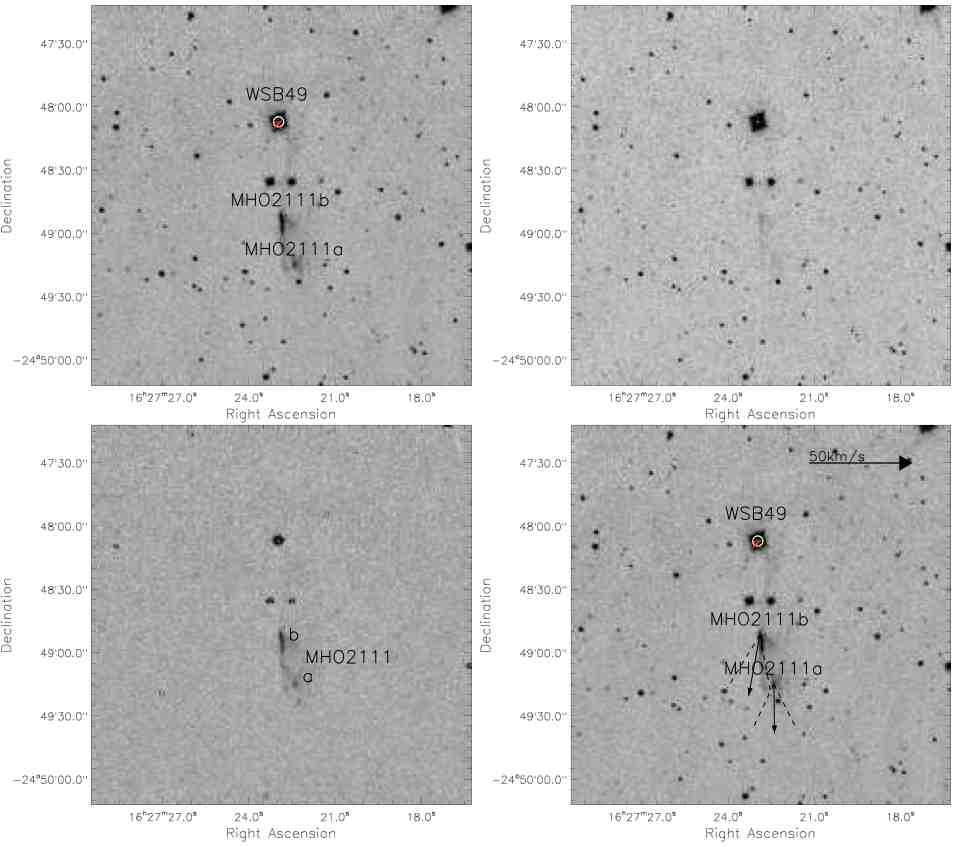}
\caption{Same as Fig.~\ref{figA1003n}, but for the region of MHO 2111.}
\label{figA0302}
\end{figure*}

\clearpage

\begin{figure*}[htb]
\centering
\includegraphics[scale=1.0]{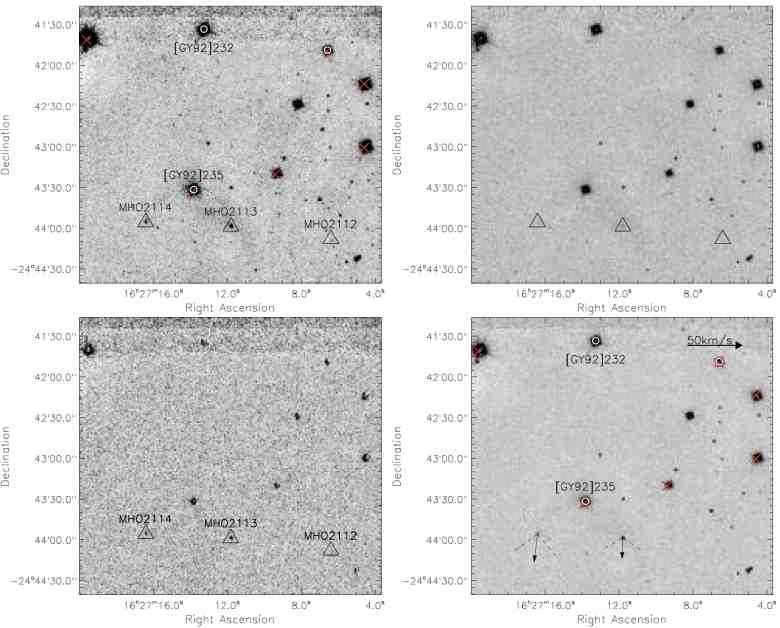}
\caption{Same as Fig.~\ref{figA1003n}, but for the region of MHO 2112-2114.}
\label{figA0402}
\end{figure*}

\begin{figure*}[htb]
\centering
\includegraphics[scale=1.0]{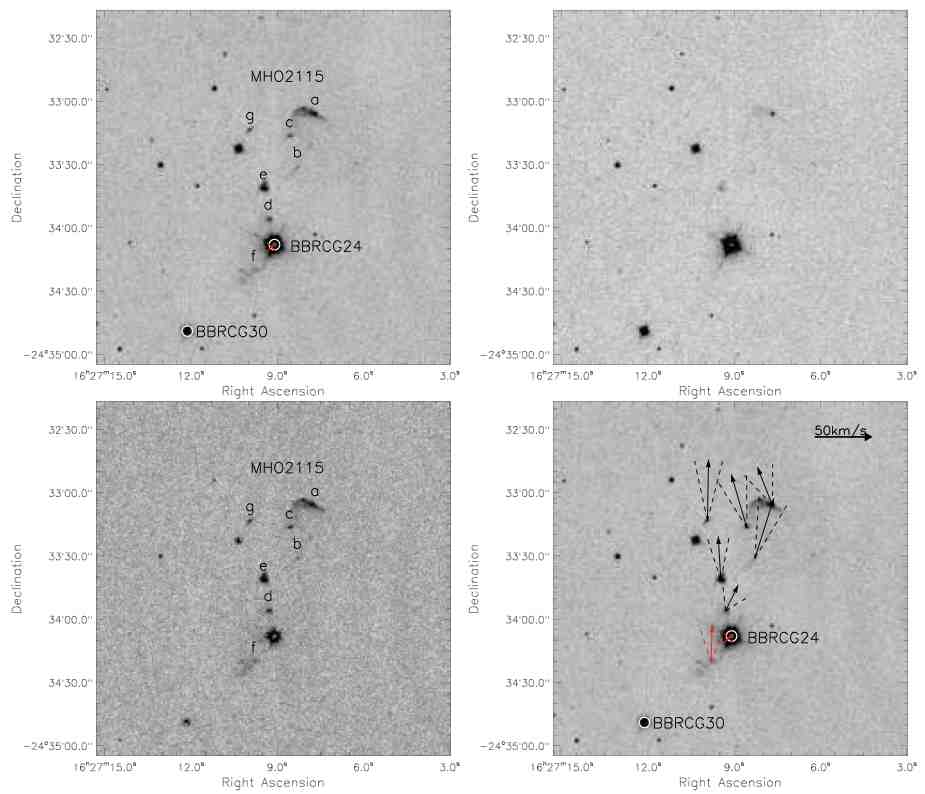}
\caption{Same as Fig.~\ref{figA1003n}, but for the region of MHO 2115.}
\label{figA0901}
\end{figure*}

\clearpage

\begin{figure*}[htb]
\centering
\includegraphics[scale=1.0]{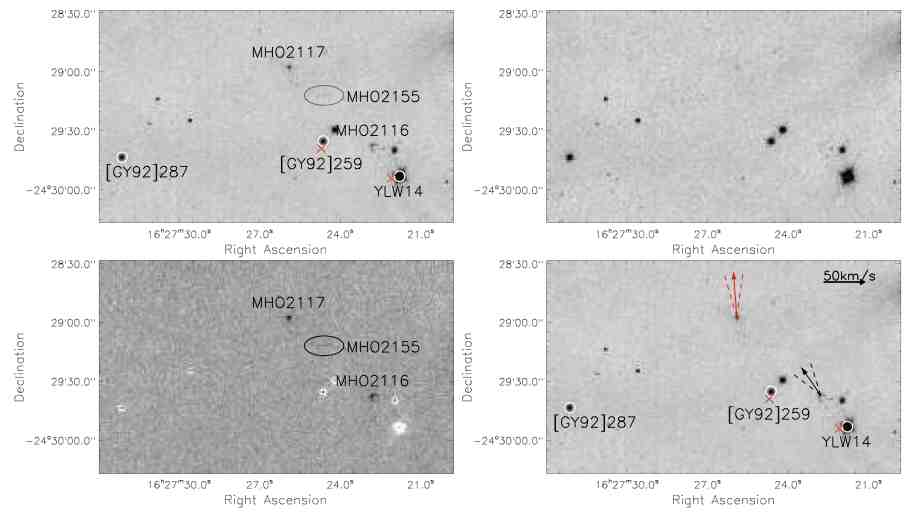}
\caption{Same as Fig.~\ref{figA1003n}, but for the region of MHO 2116 and MHO 2117.}
\label{figA0902k}
\end{figure*}

\begin{figure*}[htb]
\centering
\includegraphics[scale=1.0]{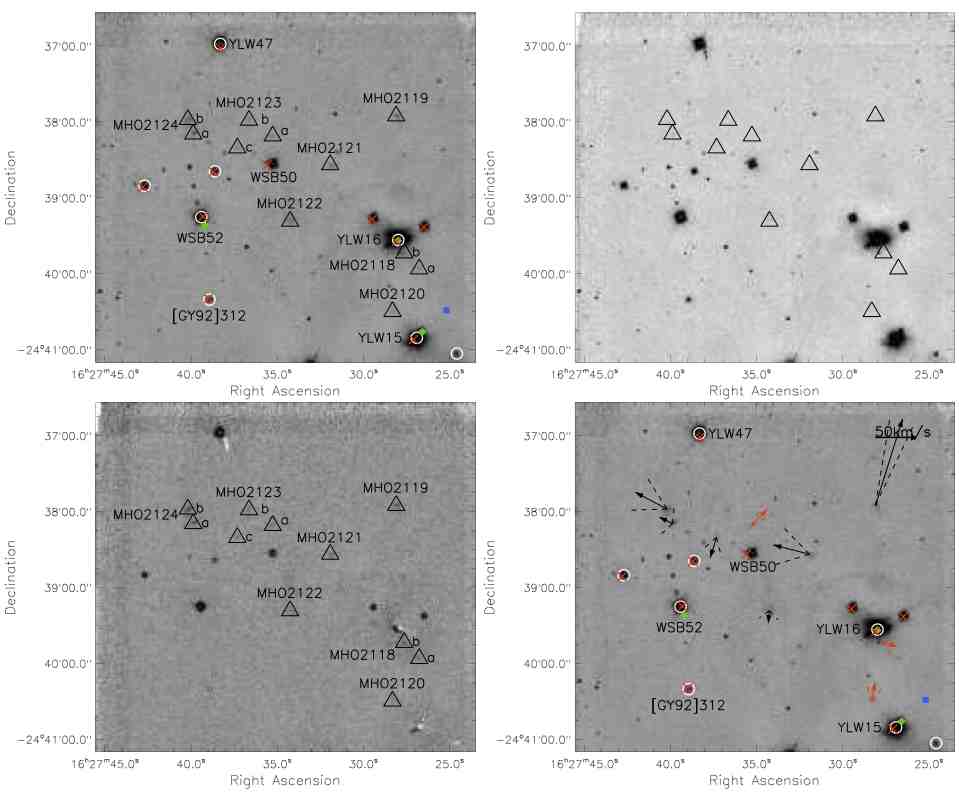}
\caption{Same as Fig.~\ref{figA1003n}, but for the region of MHO 2118-2124.}
\label{figA0403}
\end{figure*}

\begin{figure*}[htb]
\centering
\includegraphics[scale=1.0]{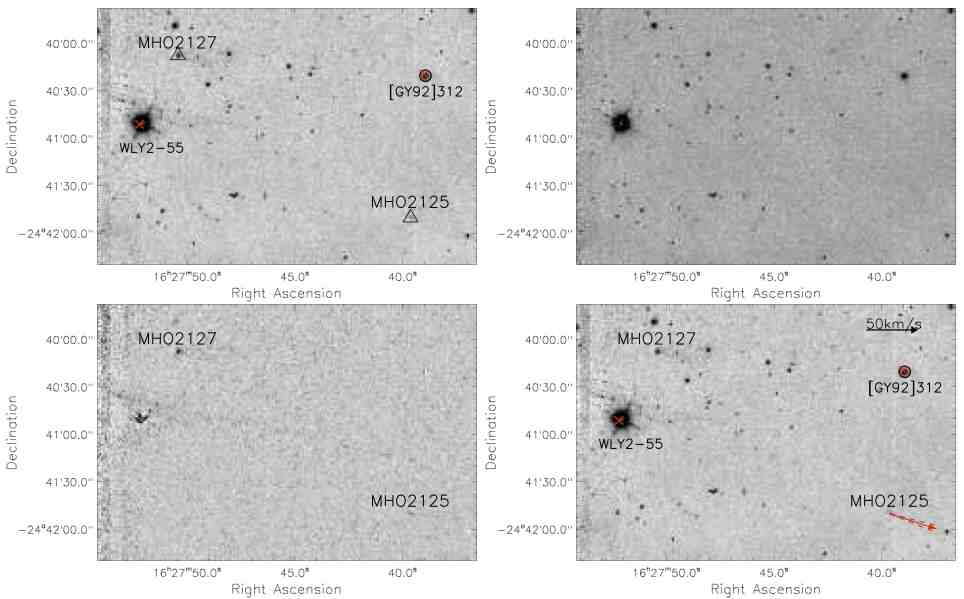}
\caption{Same as Fig.~\ref{figA1003n}, but for the region of MHO 2125 and MHO 2127.}
\label{figA0501}
\end{figure*}

\clearpage

\begin{figure*}[htb]
\centering
\includegraphics[scale=1.0]{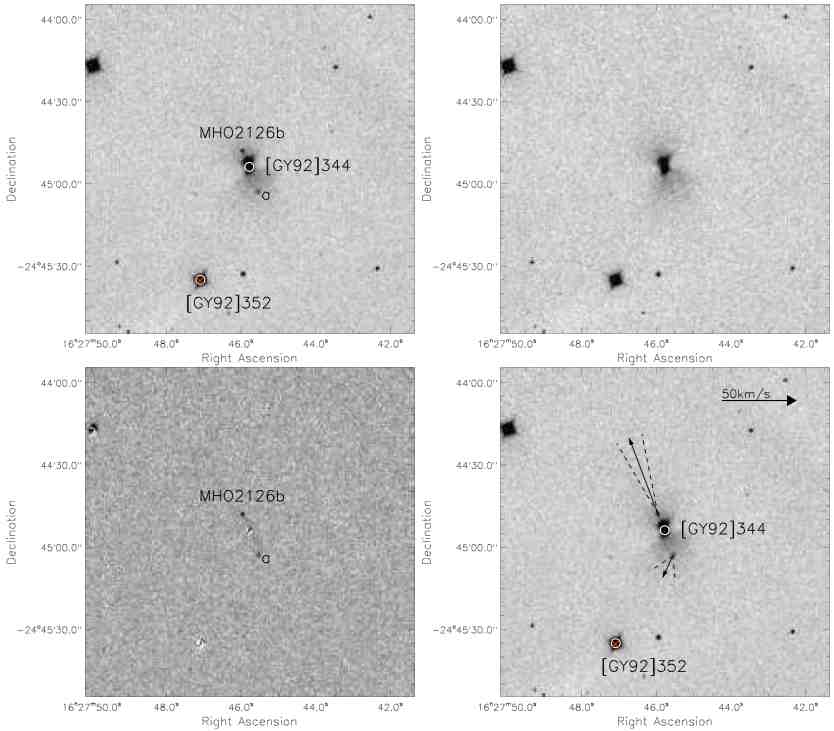}
\caption{Same as Fig.~\ref{figA1003n}, but for the region of MHO 2126.}
\label{figA0502}
\end{figure*}

\begin{figure*}[htb]
\centering
\includegraphics[scale=1.0]{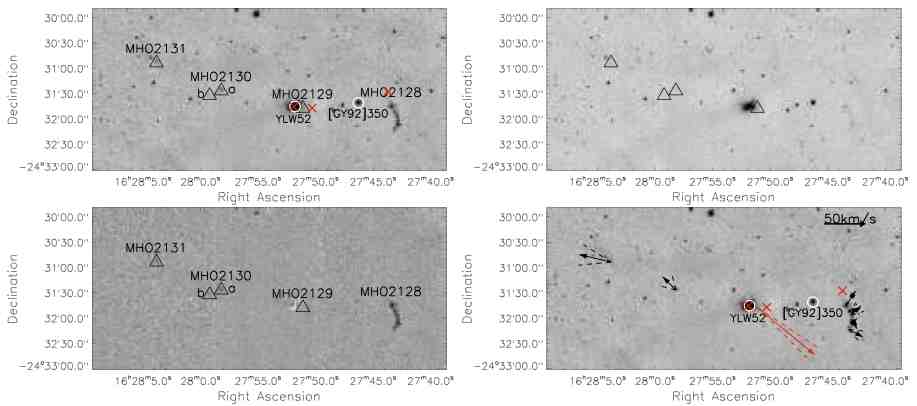}
\caption{Same as Fig.~\ref{figA1003n}, but for the region of MHO 2128-2131.}
\label{figA0801}
\end{figure*}

\begin{figure*}[htb]
\centering
\includegraphics[scale=1.0]{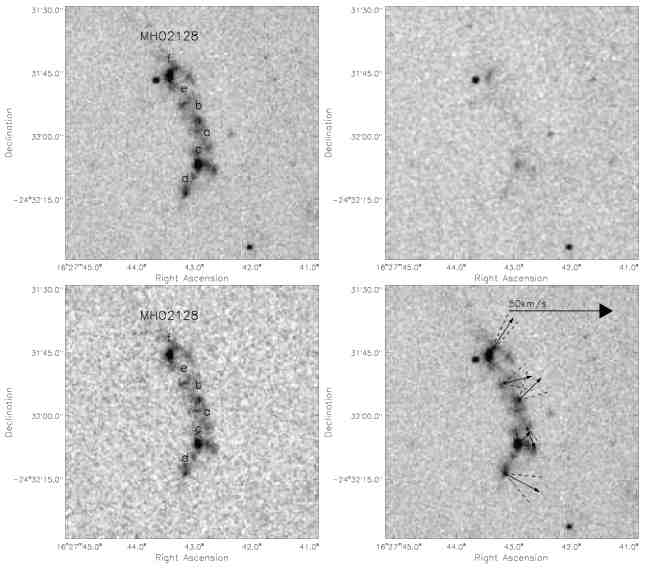}
\caption{Same as Fig.~\ref{figA1003n}, but for the region of MHO 2128.}
\label{figAp28}
\end{figure*}

\clearpage

\begin{figure*}[htb]
\centering
\includegraphics[scale=1.0]{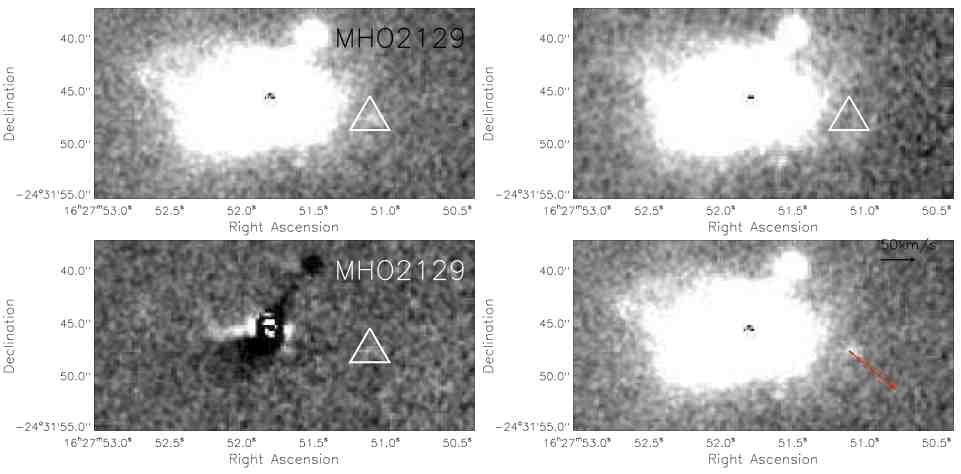}
\caption{Same as Fig.~\ref{figA1003n}, but for the region of MHO 2129. Note that backgrounds in the four panels are inverted gray scale images in order to enhance the H$_2$ feature that is marked with white triangle.}
\label{figAp29}
\end{figure*}

\begin{figure*}[htb]
\centering
\includegraphics[scale=1.0]{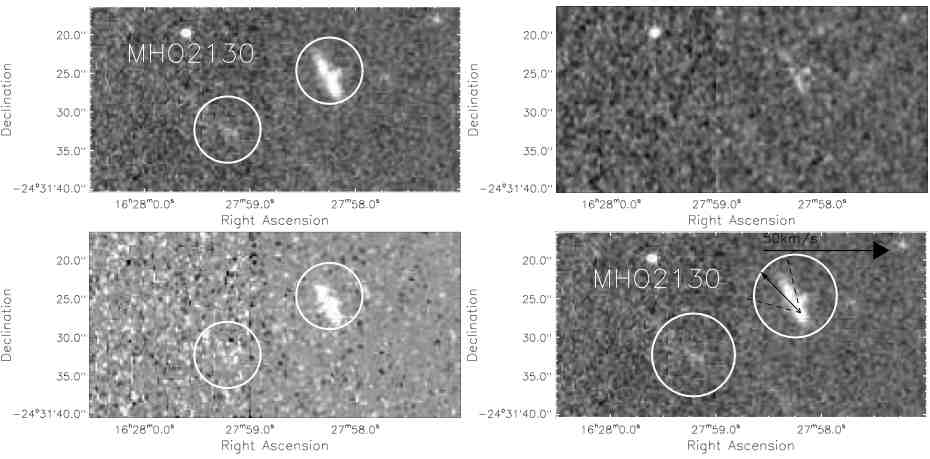}
\caption{Same as Fig.~\ref{figA1003n}, but for the region of MHO 2130. Note that backgrounds in the four panels are inverted gray scale images in order to enhance the H$_2$ features that are marked with white circles.}
\label{figAp30}
\end{figure*}

\clearpage

\begin{figure*}[htb]
\centering
\includegraphics[scale=1.0]{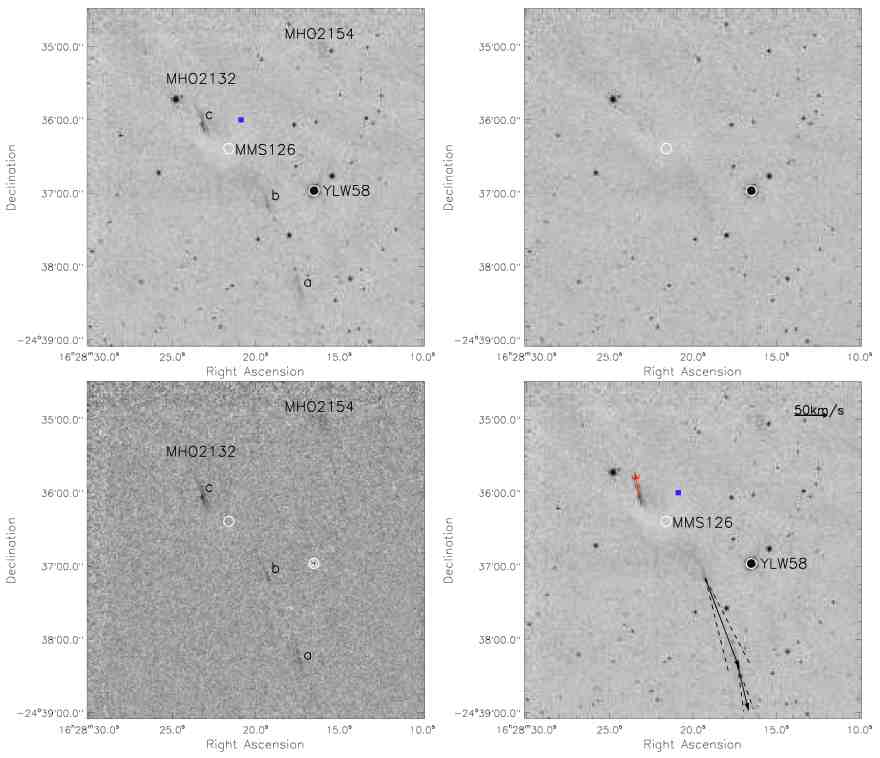}
\caption{Same as Fig.~\ref{figA1003n}, but for the region of MHO 2132 and MHO 2154.}
\label{figA0504}
\end{figure*}

\clearpage

\begin{figure*}[htb]
\centering
\includegraphics[scale=1.0]{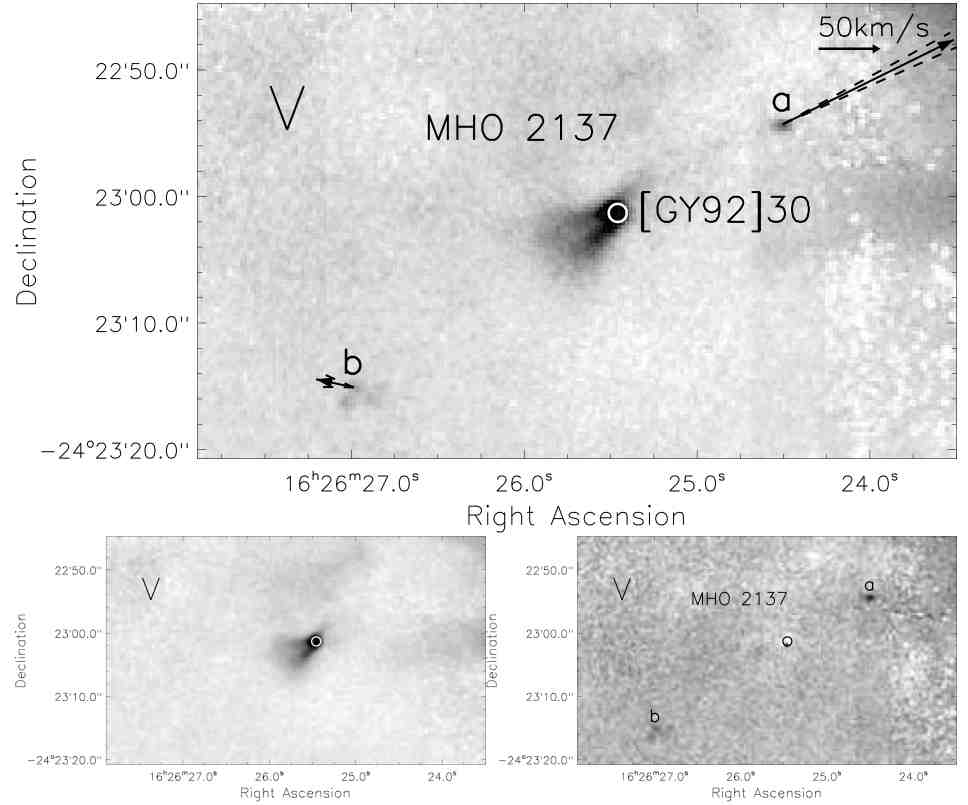}
\caption{Region of MHO 2137 and zoom-in view of box V in Fig.~\ref{figA1003main}. The \textit{top} panel shows the H$_{2}$ image with MHOs, YSOs, and PMs of MHO features being marked. The \textit{bottom-left} panel shows the Ks continuum image and the \textit{bottom-right} panel shows the continuum-subtracted H$_{2}$ image with MHO features being marked. Others are the same as Fig.~\ref{figA1003n}. }
\label{figApV}
\end{figure*}

\begin{figure*}[htb]
\centering
\includegraphics[scale=1.0]{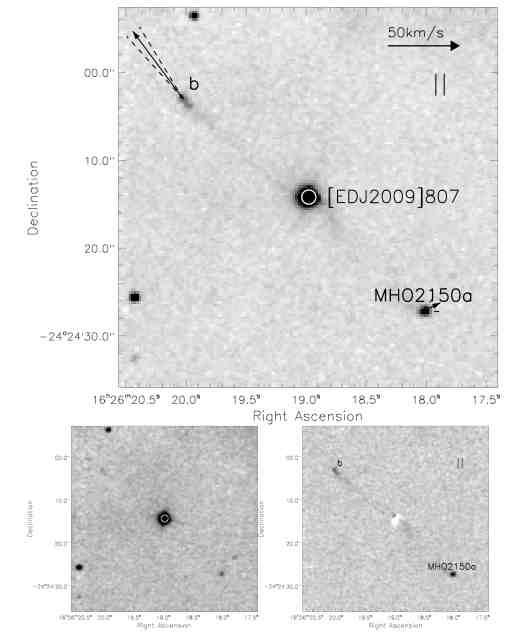}
\caption{Same as Fig.~\ref{figApV}, but for the region of MHO 2150 and box II in Fig.~\ref{figA1003main}.}
\label{figApII}
\end{figure*}

\clearpage

\begin{figure*}[htb]
\centering
\includegraphics[scale=1.0]{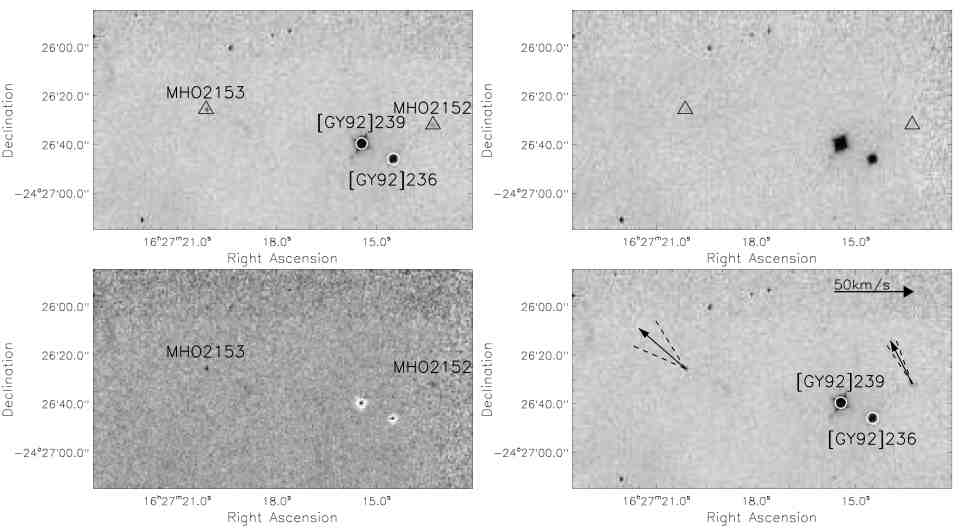}
\caption{Same as Fig.~\ref{figA1003n}, but for the region of MHO 2152 and MHO 2153.}
\label{figA0902n}
\end{figure*}

\clearpage

\end{document}